\documentclass{ieeeaccess}
\usepackage{cite}
\usepackage{amsmath,amssymb,amsfonts}
\usepackage{algorithmic}
\usepackage{graphicx}
\usepackage{textcomp}
\def\BibTeX{{\rm B\kern-.05em{\sc i\kern-.025em b}\kern-.08em
    T\kern-.1667em\lower.7ex\hbox{E}\kern-.125emX}}
\begin{document}
\history{Date of publication xxxx 00, 0000, date of current version xxxx 00, 0000.}
\doi{10.1109/ACCESS.2017.DOI}

\title{Ultra-high extinction dual-output thin-film lithium niobate intensity modulator}

\author{Sean Nelan,\authorrefmark{1,2,*} Andrew Mercante,\authorrefmark{2} Shouyuan Shi,\authorrefmark{1,2} Peng Yao,\authorrefmark{2} Eliezer Shahid,\authorrefmark{2} Benjamin Shopp,\authorrefmark{2} Cooper Hurley,\authorrefmark{1} Mathew Zablocki,\authorrefmark{2} and Dennis W. Prather\authorrefmark{1,2}}

\address[1]{School of Electrical and Computer Engineering, University of Delaware, Newark, Delaware 19716, USA}

\address[2]{Phase Sensitive Innovations, Newark, Delaware 19713, USA}

\tfootnote{This manuscript (AFRL-2022-2294) was supported in part under AFRL contract FA8650-19-C-1027. The authors gratefully acknowledge the support of UTA16-001296. The views and conclusions contained herein are those of the authors and should not be interpreted as necessarily representing the official policies or endorsements, either expressed or implied, of Air Force Research Laboratory, the Department of Defense, or the U.S. Government.}

\markboth
{Author \headeretal: Preparation of Papers for IEEE TRANSACTIONS and JOURNALS}
{Author \headeretal: Preparation of Papers for IEEE TRANSACTIONS and JOURNALS}

\corresp{Corresponding author: Sean P. Nelan (e-mail: snelan@udel.edu).}

\begin{abstract}

A low voltage, wide bandwidth compact electro-optic modulator is a key building block in the realization of tomorrow's communication and networking needs. Recent advances in the fabrication and application of thin-film lithium niobate, and its integration with photonic integrated circuits based in silicon make it an ideal platform for such a device. In this work, a high-extinction dual-output folded electro-optic Mach Zehnder modulator in the silicon nitride and thin-film lithium niobate material system is presented. This modulator has an interaction region length of 11 mm and a physical length of 7.8 mm. The device demonstrates a fiber-to-fiber loss of roughly 12 dB using on-chip fiber couplers and DC half wave voltage (V$\pi$) of less than 3.0 V, or a modulation efficiency (V$\pi\cdot$L) of 3.3 V$\cdot$cm. The device shows a 3 dB bandwidth of roughly 30 GHz. Notably, the device demonstrates a power extinction ratio over 45 dB at each output port without the use of cascaded directional couplers or additional control circuitry; roughly 31 times better than previously reported devices. Paired with a balanced photo-diode receiver, this modulator can be used in various photonic communication systems. Such a detecting scheme is compatible with complex modulation formats such as differential phase shift keying and differential quadrature phase shift keying, where a dual-output, ultra-high extinction device is fundamentally paramount to low-noise operation of the system.

\end{abstract}


\titlepgskip=-15pt

\maketitle

\section{Introduction}

The systems that support the world's communication networks are showing their age\cite{alvarado_impact_2016}. In the last decade, global data traffic has and continues to grow exponentially while existing infrastructure is becoming more costly, less efficient and less reliable as its capability is overextended\cite{alvarado_impact_2016,winzer_scaling_2017,cheng_recent_2018}. An urgency is now placed on network providers to enhance networks and meet greater demands\cite{cheng_recent_2018,yue_experimental_2019}. At the forefront of high-speed networking is the need to transmit data to networking hubs efficiently and reliably over long distances without the need for in-line amplification or repeater stations\cite{wooten_review_2000}. Existing coaxial cables are heavy, lossy and make long range data transmission difficult\cite{carey_millimeter_2021}. In turn, the shift to optical fiber-based transmission has become more desirable. Here, the optical modulator is used to overlay an electrical signal onto an optical signal for long range transmission through an optical fiber, be it to an antenna atop a radio tower, or to a networking hub hundreds of miles away\cite{wooten_review_2000}. The optical modulator is an integral part of active and passive millimeter wave imaging systems, modern telecommunications networks and data communication, and is widely used in on-chip RF photonic devices, frequency comb generation, on chip signal splitting, sensing, and quantum photonics\cite{beardell_rf-photonic_2021,nelan_compact_2022,liu_low_2021,chen_high_2022,zhang_high_2016,alvarado_impact_2016,yue_experimental_2019,zhang_integrated_2021,he_high-performance_2019,ahmed_high-efficiency_2020,mercante_thin_2018,rao_compact_2018,reed_silicon_2010,weigel_bonded_2018,horst_cascaded_2013,huo_diamond_2016,wang_integrated_2018,alloatti_100_2014,haffner_all-plasmonic_2015,coward_photonic_1993}. 

If an optical modulator is to be integrated with existing systems, it should maintain a small device footprint, feature a wide RF bandwidth and remain environmentally stable for reliable operation in any environment\cite{mercante_thin_2018, ahmed_subvolt_2020, ahmed_high-efficiency_2020, wang_integrated_2018, rao_compact_2018}. Moreover, the ideal optical modulator boasts a high (>35dB) extinction ratio\cite{ahmed_high-efficiency_2020,nelan_compact_2022,jin_high-extinction_2019,chen_high_2022,wang_thin-film_2022}. At first glance, Si-based free carrier plasma dispersion-based modulators would seem the ideal candidate, but their low cost and excellent scalability is overshadowed by poor extinction ratio and bandwidth; an inherent limitation of a dispersion-based modulation scheme\cite{reed_silicon_2010, subbaraman_recent_2015, azadeh_low_2015}. Si-based modulation is further blighted by high intrinsic absorption loss, narrow transmission spectrum and dopant diffusion at high temperatures\cite{ahmed_high-efficiency_2020, reed_silicon_2010}. Finally, because Si is a centrosymmetric material, it demonstrates a high third-order non-linearity and a low second-order non-linearity, which excludes it from efficient operation at high optical powers\cite{dinu_third-order_2003}. While the integration of organic electro-optic polymers with the Si platform has been shown to mitigate some of its inherent drawbacks, the organic nature of the platform does not lend itself to long-term environmental stability\cite{zhang_high_2016, min-cheol_oh_recent_2001, zhang_high_2016, alloatti_100_2014}.

Si-based free carrier plasma dispersion-based modulation platform is then ruled out in favor of lithium niobate-based modulation, where the ideal platform for a low-voltage, high-bandwidth and environmentally-stable modulator should show a strong electro-optic (Pockels) effect, and a linear response to an applied modulation voltage irrespective of the optical power and wavelength\cite{zhang_integrated_2021, wang_integrated_2018,nelan_compact_2022,ahmed_subvolt_2020,mercante_thin_2018,rao_compact_2018}. Lithium niobate (LiNbO$_\mathrm3$) brings an extremely strong second-order non-linearity ($\chi^{(2)}$), a strong linear electro-optic effect, low optical absorption across a wide range of wavelengths, pure phase modulation, zero chirping and exceptional stability at high temperatures\cite{nelan_compact_2022,ahmed_subvolt_2020,mercante_thin_2018}. LiNbO$_\mathrm{3}$, with its trigonal crystal system, lacks inversion symmetry and boasts a third-order non-linearity ($\chi^{(3)}$) which is three magnitudes lower than Si\cite{dinu_third-order_2003,nelan_compact_2022}.

Commercial, or bulk LiNbO$_\mathrm{3}$ modulators use titanium (Ti)-diffused waveguides to carry the optical mode, but suffer poor optical confinement ($\Delta$n < 0.02) which results in a large optical mode size\cite{nelan_compact_2022,ahmed_subvolt_2020}. This requires metal electrodes to be placed far from the optical waveguide to avoid metal absorption loss, and results in large bending radii. Consequentially, the V$\pi$ and the size of the device is increased. A typical commercial LiNbO$_\mathrm{3}$ modulator brings a V$\pi\cdot$L of 12-32 V$\cdot$cm\cite{ahmed_subvolt_2020,wang_integrated_2018,zhang_integrated_2021}.

Recently, advances in crystal ion sliced (CIS) films of LiNbO$_\mathrm{3}$ on insulator (TFLNOI), which guide optical modes almost 20 times smaller than their bulk-LiNbO$_\mathrm{3}$ counterparts have emerged as an answer to some of these issues \cite{nelan_compact_2022,wang_nanophotonic_2018,rao_compact_2018,ahmed_subvolt_2020,mercante_thin_2018,rao_high-performance_2016}. Now, strip-loaded waveguides can be used to tightly confine the optical mode, allowing smaller electrode gaps, decreased V$\pi$, tighter bending radii and photonic integrated circuit (PIC) compatibility. In previous work, ultra-wide bandwidth, high efficiency and low loss modulators have been demonstrated in the TFLNOI material system with ridge-etched and strip-loaded waveguides\cite{nelan_compact_2022,ahmed_subvolt_2020,ahmed_high-efficiency_2020,mercante_thin_2018}. To facilitate the use of a mode-transition coupler at the input and output of the device, silicon nitride (SiN$_\mathrm{x}$) was again chosen as the ideal candidate to create a strip loaded waveguide in the TFLNOI platform. SiN$_\mathrm{x}$ has low propagation loss across telecom optical wavelengths, high power handling capability and a similar optical index to LiNbO$_\mathrm3$\cite{huffman_integrated_2018,wooten_review_2000,jin_linbo_2016,stocchi_mid-infrared_2019,kaloyeros_reviewsilicon_2017}. SiN$_\mathrm{x}$ also features a low third-order non-linearity, small thermo-optic coefficient, and can be deposited using plasma enhanced chemical vapor deposition (PECVD)\cite{huffman_integrated_2018,wooten_review_2000,jin_linbo_2016,stocchi_mid-infrared_2019,kaloyeros_reviewsilicon_2017}.

One must also consider the packaging and size of the device. A conventional push-pull electro-optic modulator utilizes a straight waveguide before and after the interaction region, where light enters one side of the device and exits the opposite side, requiring fiber coupling to both sides of the device simultaneously \cite{ahmed_subvolt_2020,mercante_thin_2018,wang_nanophotonic_2018}. To reduce the size, weight, power and cost (SWaP-C) of the system, the device can be "folded." Here, a 180$^{\circ}$ bend is introduced to the waveguides and electrodes to direct the signal back towards the input. Now, the length of the device is halved with no adverse affect on modulation efficiency or bandwidth\cite{nelan_compact_2022,sun_folded_2021,hu_folded_2021}. The modulator can then be used with a single, easily packaged v-groove array (VGA), can be folded multiple times with little impact to the overall footprint, and only requires the polishing and preparation of one end-facet\cite{nelan_compact_2022,hu_folded_2021}. However, this layout does introduce some challenges to the design and fabrication of the modulator. If a standard Mach-Zehnder Modulator (MZM) were to be bent back on itself such that the interaction regions before and after the bend were the same length, the latter region would reverse any modulation imparted by the first, because the polarity of the modulating electric field interacting with the mode inside the LiNbO$_\mathrm{3}$ is now the inverse of what it was during the first half of the modulator\cite{nelan_compact_2022,hu_folded_2021,sun_folded_2021}. Moreover, the optical path lengths of the inner and outer interaction regions would be mismatched, and the device would show a frequency-dependent response.

\Figure[h](topskip=0pt, botskip=0pt, midskip=0pt)[width=0.6\textwidth]{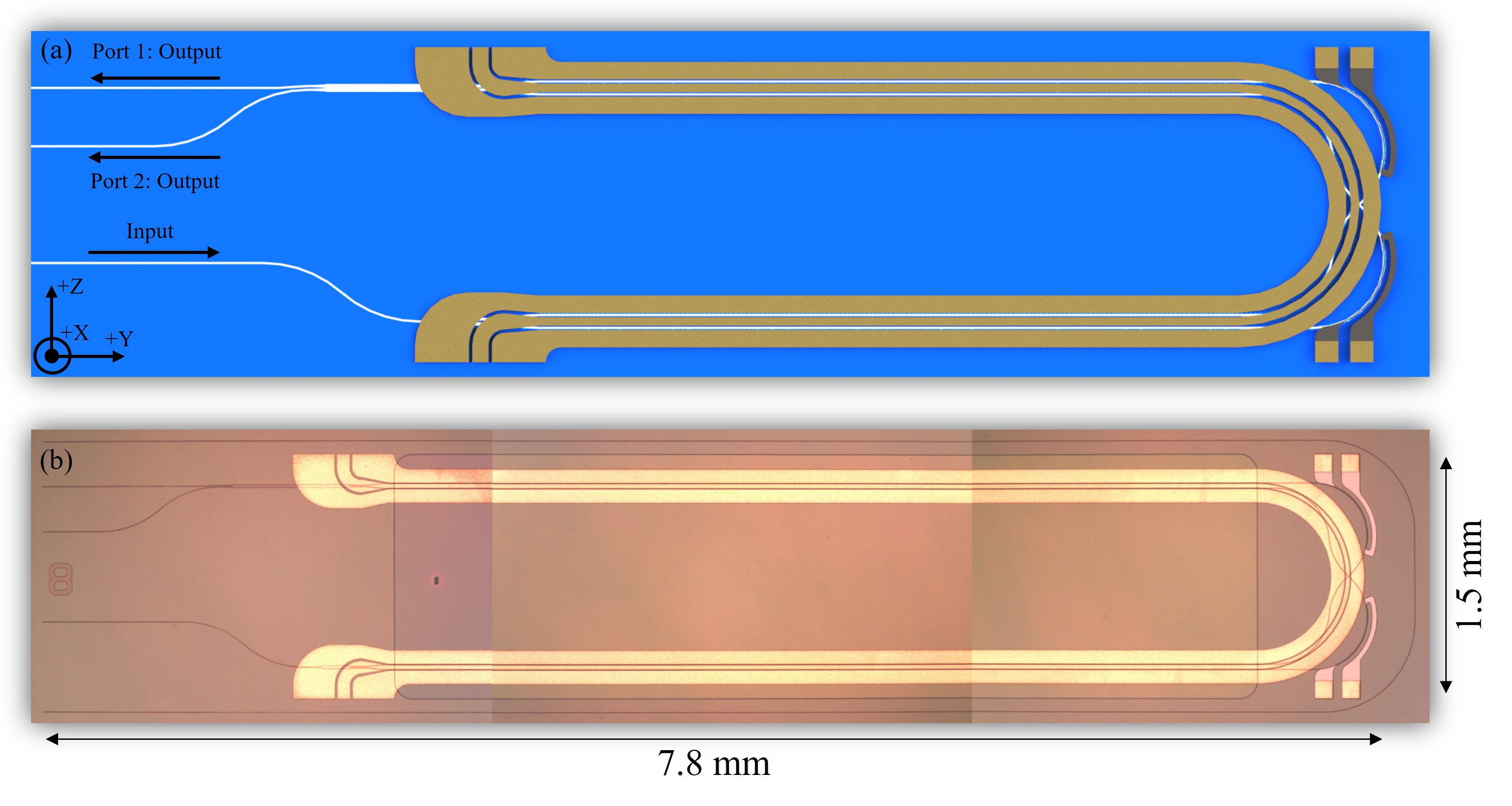}
{(a) Aerial view of a rendering of the folded modulator with waveguide crossing showing crystal axis and device port direction.
(b) Aerial view of the fabricated folded modulator with waveguide crossing. }
\label{fig:solidworks}

In previous work, a waveguide crossing was used to ensure both sections of the modulator, before and after the fold, contribute to the total modulation of the device\cite{nelan_compact_2022}. A crossing is again used in this device to minimize the overall footprint. The optical path length of both arms, and the optical vs. electrical path length is then matched to allow operation without any wavelength-dependent effects or RF and optical field separation. A symmetric 2x2 multi-mode interference (MMI) splitter is placed at the output of the device to yield two outputs with intensities inversely proportional to each other. Paired with a balanced photo-diode (PD) receiver, a dual-output MZM can be used in many RF photonic link systems that require an extremely low noise floor and spurious-free dynamic range (SFDR) performance. This is vital to communication systems and enables heavy, lossy and space-consuming cables to be replaced with light, almost lossless and small volume optical fibers \cite{carey_millimeter_2021}. A dual-output MZM/balanced PD receiver pair is capable of detecting not only the signal amplitude, but also the phase information. As a result, such detecting scheme is compatible with complex modulation formats such as differential phase shift keying (DPSK) and differential quadrature phase shift keying (DQPSK)\cite{zhang_using_2019}.

Through careful design of input and output splitters, and the use of symmetric MMI splitters, an extinction ratio over 45 dB is achieved at each output port. This design does not require cascaded directional couplers and additional control circuitry to control splitting imbalance, but the optical loss in both waveguides in the interaction region of the modulator must be closely matched through careful fabrication to achieve the high extinction ratio \cite{jin_high-extinction_2019}. This will be discussed later in the work. The measurement shows repeatability across multiple devices.

In this work, a high extinction ratio hybrid LiNbO$_\mathrm{3}$-SiN$_\mathrm{x}$ electro-optic MZM with two equal and opposite outputs is designed, fabricated and characterized. Fiber couplers are used at the input and output to lower the total insertion loss to 12 dB. To the best of our knowledge, this is the first time a high extinction ratio, folded, LiNbO$_\mathrm{3}$-SiN$_\mathrm{x}$ dual-output MZM with a waveguide crossing has been demonstrated in this material platform. The device presented shows a measured DC-V$\pi$ of 3.0 V with an 11 mm interaction region, and an extinction ratio of over 45 dB.

\section{Design and simulation}

\subsection{Device layout}

The 3D aerial view of the folded electro-optic modulator, including thermal biasing heaters is shown in Fig. \ref{fig:solidworks}. (a) and (b). The electrode contains one interaction region before and after the folded region, each 5.5 mm long. The input and output ports are defined in Fig. \ref{fig:solidworks}. (a). The material stack, from top to bottom, is comprised of a 450 nm buffer layer of PECVD SiO$_\mathrm{2}$ (T$_\mathrm{BUF}$), 100 nm of PECVD SiN$_\mathrm{x}$ deposited on 300 nm X-cut thin-film LiNbO$_\mathrm{3}$, which is bonded to a 4.7 $\mu$m SiO$_\mathrm{2}$ layer (T$_\mathrm{BBUF}$), atop a 500 $\mu$m Si handle. To efficiently guide the first order transverse electric (TE) mode, the width (W$_\mathrm{SiN}$) and thickness (T$_\mathrm{SiN}$) of the strip-loaded SiN$_\mathrm{x}$ waveguide is chosen to be 2 $\mu$m and 100 nm, respectively. The simulated fundamental TE mode in the interaction region of the device can be seen in Fig. \ref{fig:modefield}. (a). The waveguide cross section in the interaction region and material stack are defined in Fig. \ref{fig:modefield}. (c) and (d), respectively. The device uses a fiber coupler at the end facet to minimize free-space coupling loss caused by the mode field diameter (MFD) mismatch between the lensed fiber and the waveguide. The mode field in the fiber coupler can be seen in Fig. \ref{fig:modefield}. (b). The optical mode propagates in the Y-crystal direction, while the electrodes are placed such that the RF field is polarized in the +/- Z-crystal direction, where the maximum r$_\mathrm{33}$ = 31 pm/V coefficient is achieved for the TE optical mode \cite{ahmed_subvolt_2020,jin_linbo_2016,ren_integrated_2019,wooten_review_2000,nelan_compact_2022}. The layout and design of the folded modulator with a waveguide crossing are further discussed in previous work where a standard single output folded MZM is reported\cite{nelan_compact_2022}.

A 2x2 MMI splitter that recombines the light to produce an intensity-modulated signal is deployed on the output of the device. The symmetric 2x2 multi-mode interference (MMI) splitter yields two outputs which are inversely proportional to each other. When the device is biased such that one output is at it's peak intensity, the other output is at it's lowest, or null intensity. The device can then be paired with a balanced PD receiver to eliminate any common-mode noise between both outputs\cite{jin_balanced_2015}.

\Figure[h](topskip=0pt, botskip=0pt, midskip=0pt)[width=0.9\columnwidth]{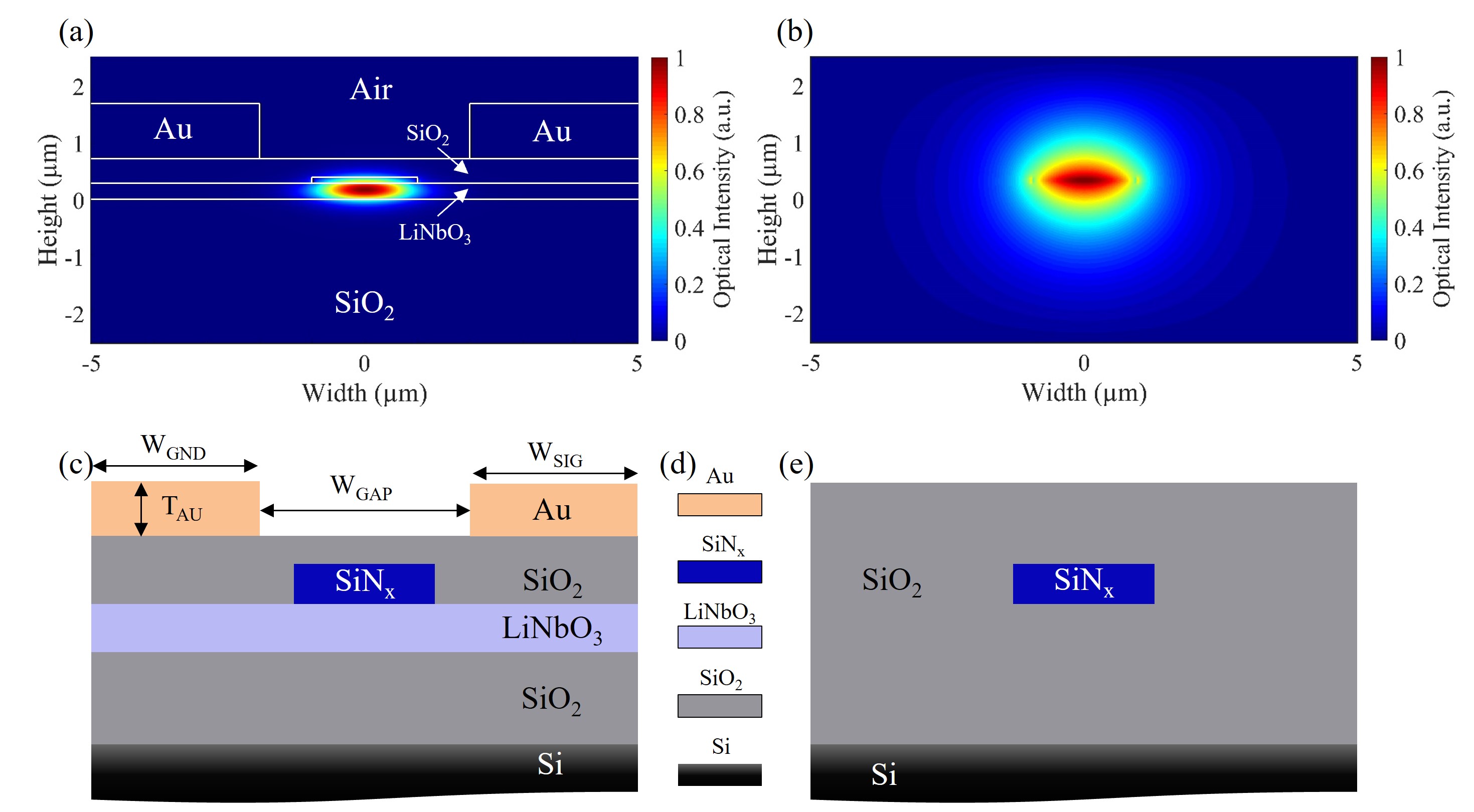}
{(a) Simulated mode field of the waveguide in the interaction region. Simulations done in Lumerical MODE Solver. 
(b) Simulated mode field of the waveguide at the end facet. Simulations done in Lumerical MODE Solver. 
(c) Cross-sectional view of a single arm of the material structure in the interaction region. Dimensions are labeled. 
(d) Material definitions.
(e) Cross-sectional view of the material structure at the device end facet.}
\label{fig:modefield}

\subsection{2x2 MMI splitter design}

An MMI splitter and combiner are used as 3 dB couplers to both split and recombine light before and after the interaction region of the device, respectively. Directional couplers and Y-splitters which require an extremely small gap between waveguides are sensitive to fabrication errors\cite{jin_high-extinction_2019}. The MMI offers greater fabrication tolerance, and in the case of the 1x2 MMI, the symmetric design does not require precise, sub-micron accuracy of device length to maintain an equal splitting ratio\cite{ahmed_subvolt_2020,khalil_two-dimensional_2004}. The design, simulation and fabrication of the 1x2 MMI splitter is reported in previous work\cite{nelan_compact_2022}.

To achieve an inversely-proportional, balanced output from the MZM, a 2x2 MMI is used at the output of the device following the interaction region. The 2x2 MMI operates based on the self-imaging principle, where any input field is reconstructed periodically after a certain propagation distance\cite{khalil_two-dimensional_2004}. With this in mind, the length (L$_\mathrm{MMI}$) of the 2x2 MMI must be designed such that the input field from either port is reconstructed into two output fields at the end of the device. The width of the MMI is then set to 15 $\mu$m to avoid evanescent coupling of the output fields after the multimode region. The mode separation distance (S$_\mathrm{MMI}$) is 8 $\mu$m between both ports of the 2x2 MMI.

\Figure[h](topskip=0pt, botskip=0pt, midskip=0pt)[width=0.9\columnwidth]{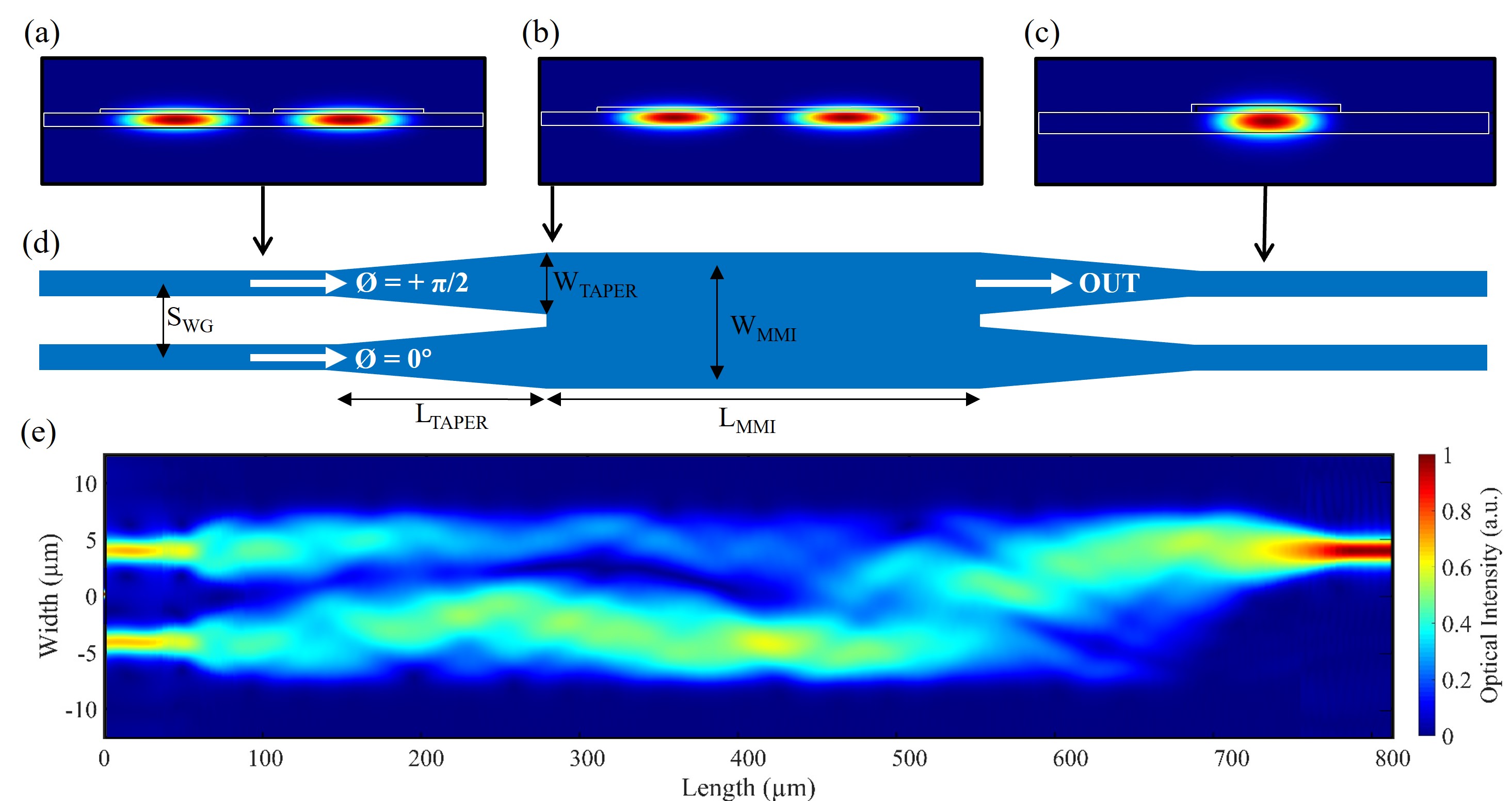}
{(a) Mode field cross section in the input waveguides of the 2x2 MMI when both input ports are illuminated with a $\pi$/2 phase difference.
(b) Mode field cross section at the input of the 2x2 MMI when both input ports are illuminated with a $\pi$/2 phase difference.
(c) Mode field cross section at the output of the 2x2 MMI when both input ports are illuminated with a $\pi$/2 phase difference.
(d) Aerial rendering of the 2x2 MMI. Dimensions and port propagation directions are labeled.
(e) Mode cross section in the +X crystal axis of the 2x2 MMI when both input ports are illuminated with a $\pi$/2 phase difference. Propagation occurs from left to right.}
\label{fig:mmi}

The initial length of the multimode waveguide region is estimated using the 2D mathematical approximation given by \cite{ahmed_subvolt_2020,khalil_two-dimensional_2004,bachmann_general_1994}:
\begin{equation}
L_\mathrm{MMI} = \frac{4n_\mathrm{r} w^{2}_\mathrm{e}}{\mathrm{N\lambda_{0}}}.
\end{equation}
Here, $n_r$ is the effective index of the multimode region, $w_e$ is the width of the multimode region (W$_\mathrm{MMI}$), N is the number of self-images (inputs and outputs), and $\lambda_0$ is the input wavelength in vacuum. The $n_\mathrm{eff}$ of the lowest order TE modes in the MMI region are found to be 2.128018, and 2.129952, respectively, using Lumerical's MODE solver. This results in a multimode region length of 618 $\mu$m. For optimization of the initial design parameters, Lumerical’s MODE simulation software is used. With W$_\mathrm{MMI}$ = 15 $\mu$m and S$_\mathrm{MMI}$ = 8 $\mu$m, the final length of the structure is found to be L$_\mathrm{MMI}$ = 640 $\mu$m, which is very close to the length given by the self-imaging method. For further reduction of the mode-index mismatch between the MMI region and the input waveguide, linear tapers of length (L$_\mathrm{TPR}$) 52.5 $\mu$m and width (W$_\mathrm{TPR}$) 7 $\mu$m are introduced at the input and output section, shown in Fig. \ref{fig:mmi}. (d). The introduction of a taper increases the range of MMI lengths over which it continues to split the input mode equally among the two outputs as well as the bandwidth of the MMI, which is wavelength-dependent as the 3 dB coupling length depends on $\frac{1}{\lambda}$\cite{nelan_compact_2022,ahmed_subvolt_2020,wang_proposal_2014}. Using Lumerical's FDTD simulation software, the optimized length of the MMI with tapered input and output ports is found to be L$_\mathrm{MMI}$ = 636 $\mu$m. This is near the length given by the self-imaging equation, but the introduction of linear tapers has an effect on the overall length of the cavity to properly reconstruct the image at the output waveguides, and slightly reduces the multimode region length. Because both the 1x2 and the 2x2 MMI are symmetrically fed when both input ports are illuminated, the input power should be equally split among the output arms barring any fabrication intolerance. This will be confirmed later in the manuscript through measuring the extinction ratio at each port. An electric field profile for the designed two-fold self-image MMI is shown in Fig. \ref{fig:mmi}. (e).

\subsection{Traveling wave electrode design}

\Figure[h](topskip=0pt, botskip=0pt, midskip=0pt)[width=0.8\columnwidth]{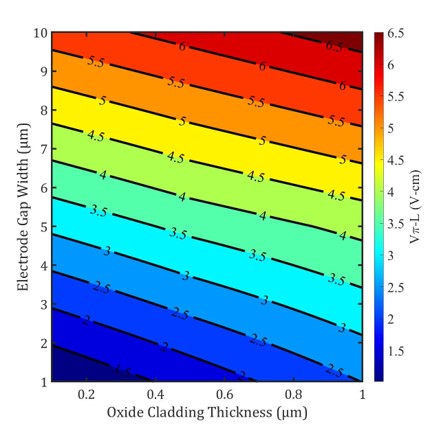}
{The effect of electrode gap width (W$_\mathrm{GAP}$) and oxide cladding thickness on V$\pi\cdot$L. }
\label{fig:heatmap}

The ground-signal-ground (GSG) co-planar waveguide (CPW) electrodes must maintain a 50 $\Omega$ impedance at the high-frequency operation limit of the device with minimal RF absorption or conduction loss to minimize reflections in the feed network. The dimensions of the GSG electrodes are defined in Fig. \ref{fig:modefield}. (c). The electrode gap (W$_\mathrm{GAP}$) must be kept as small as possible to maximize RF and optical field overlap, contributing to low-voltage operation, but must not contribute to excess optical absorption loss from the TE mode interacting with the Au electrode\cite{nelan_compact_2022,mercante_thin_2018,ahmed_subvolt_2020,wang_thin-film_2022}. In Fig. \ref{fig:heatmap}, the simulated V$\pi\cdot$L is plotted against W$_\mathrm{GAP}$ and T$_\mathrm{BUF}$. To achieve a V$\pi\cdot$L of roughly 3.3 V$\cdot$cm, or a V$\pi$ of roughly 3 V with our 11 mm interaction region, W$_\mathrm{GAP}$ is set to 4 $\mu$m and T$_\mathrm{BUF}$ is set to 450 nm. This yields an additional optical absorption loss of roughly .08 dB/cm. The width of the ground electrodes (W$_\mathrm{GND}$) is set to 75 $\mu$m, and the height of the electroplated Au electrodes (T$_\mathrm{ELEC}$) is 1 $\mu$m. The optimal electrode signal width (W$_\mathrm{SIG}$) is found to be 24.5 $\mu$m using Ansys High Frequency Structure Simulator (HFSS), in a process described in previous work \cite{nelan_compact_2022}. An index matching fluid (UV15) is applied to the GSG electrodes in the interaction region and fold to more closely match the RF and optical group velocities and improve modulation bandwidth\cite{ibarra_fuste_bandwidthlength_2013,nelan_compact_2022,mercante_thin_2018,ahmed_high-efficiency_2020}.

\Figure[h](topskip=0pt, botskip=0pt, midskip=0pt)[width=0.7\textwidth]{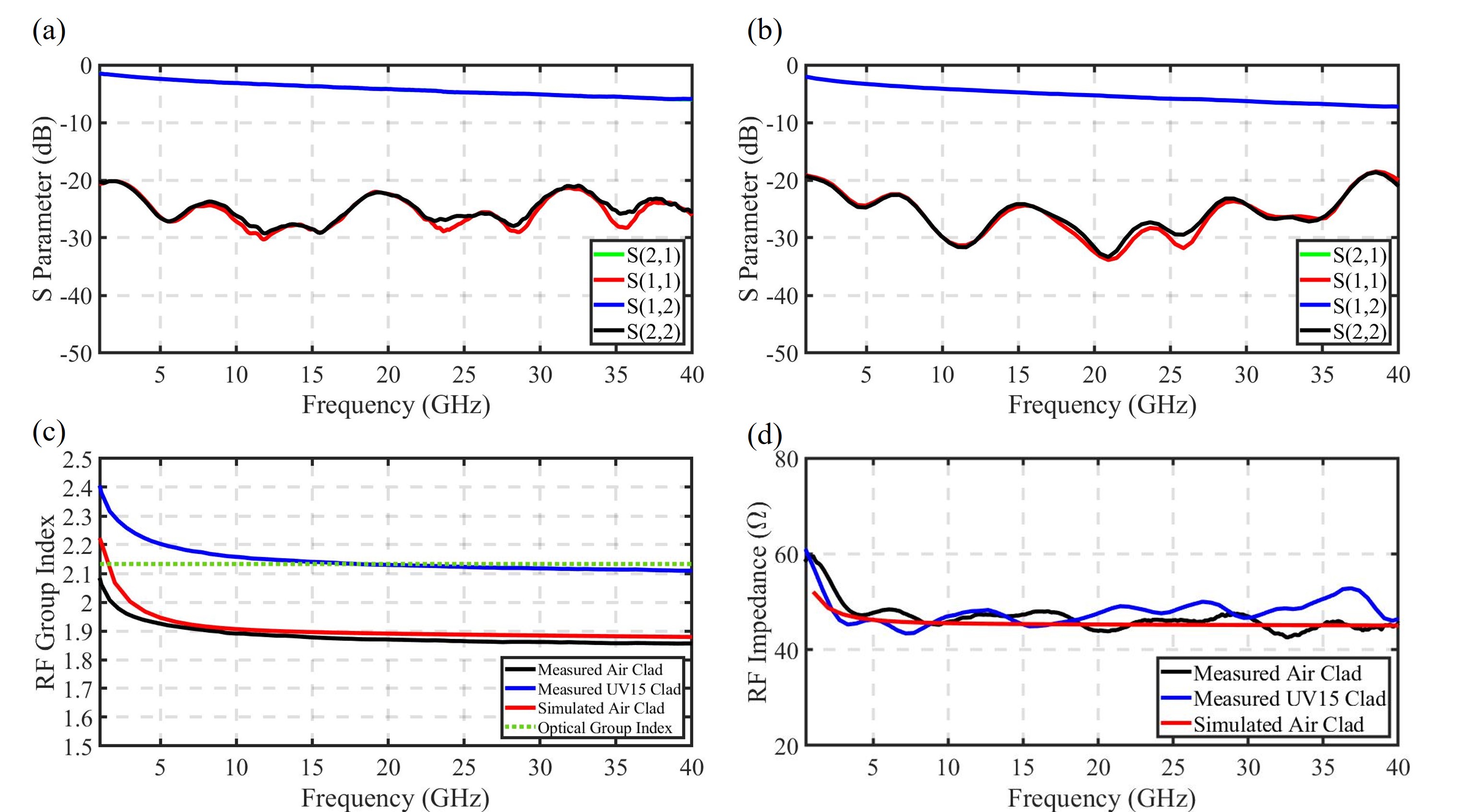}
{(a) Measured S-parameters of the air-clad CPW GSG electrodes from 0 to 40 GHz. Extracted from the real and imaginary measured S-parameter matrix.
(b) Measured S-parameters of the UV15-clad CPW GSG electrodes from 0 to 40 GHz. Extracted from the real and imaginary measured S-parameter matrix.
(c) RF group index of the GSG electrodes from 0 to 40 GHz. Extracted from the real and imaginary S(2,1) parameters.
(d) RF impedance of the GSG electrodes from 0 to 40 GHz. Extracted from the real and imaginary S(1,1) parameters.}
\label{fig:electrodes}

The high frequency electrical response of the GSG electrodes is characterized through a 2-port measurement using a Keysight PNA-X network analyzer. A 50 $\Omega$ GSG probe is connected to each end of the GSG electrodes, and a frequency sweep performed from 0 to 40 GHz. The measured air-clad and UV15-clad S-parameters are plotted in Fig. \ref{fig:electrodes} (a) and (b), respectively. At 40 GHz, S(2,1) is observed to be roughly -3 dB and -4 dB at the 40 GHz target frequency for the air-clad and UV15-clad electrode, respectively. There is a periodic resonance seen in S(1,1) and S(2,2), which is caused by constructive and destructive interference from reflections at the probe points as the frequency is swept.  The RF impedance with and without index matching fluid is plotted in Fig. \ref{fig:electrodes}. (d). The observed resonance of the electrode is seen in the measured impedance. The RF group index ($n_\mathrm{RF}$) with and without index matching fluid is plotted alongside the optical group index ($n_\mathrm{Og}$) in Fig. \ref{fig:electrodes} (c). The $n_\mathrm{Og}$ of the fundamental TE mode is found to be roughly 2.14, while the group index of the RF mode in the air-clad and UV15-clad electrodes ($n_\mathrm{RF}$) is roughly 1.85 and 2.10, respectively. The index mismatch between the optical and RF mode is then 0.04 with index matching fluid, with the RF mode propagating faster than the optical mode. 
\section{Fabrication}

\Figure[h](topskip=0pt, botskip=0pt, midskip=0pt)[width=0.9\columnwidth]{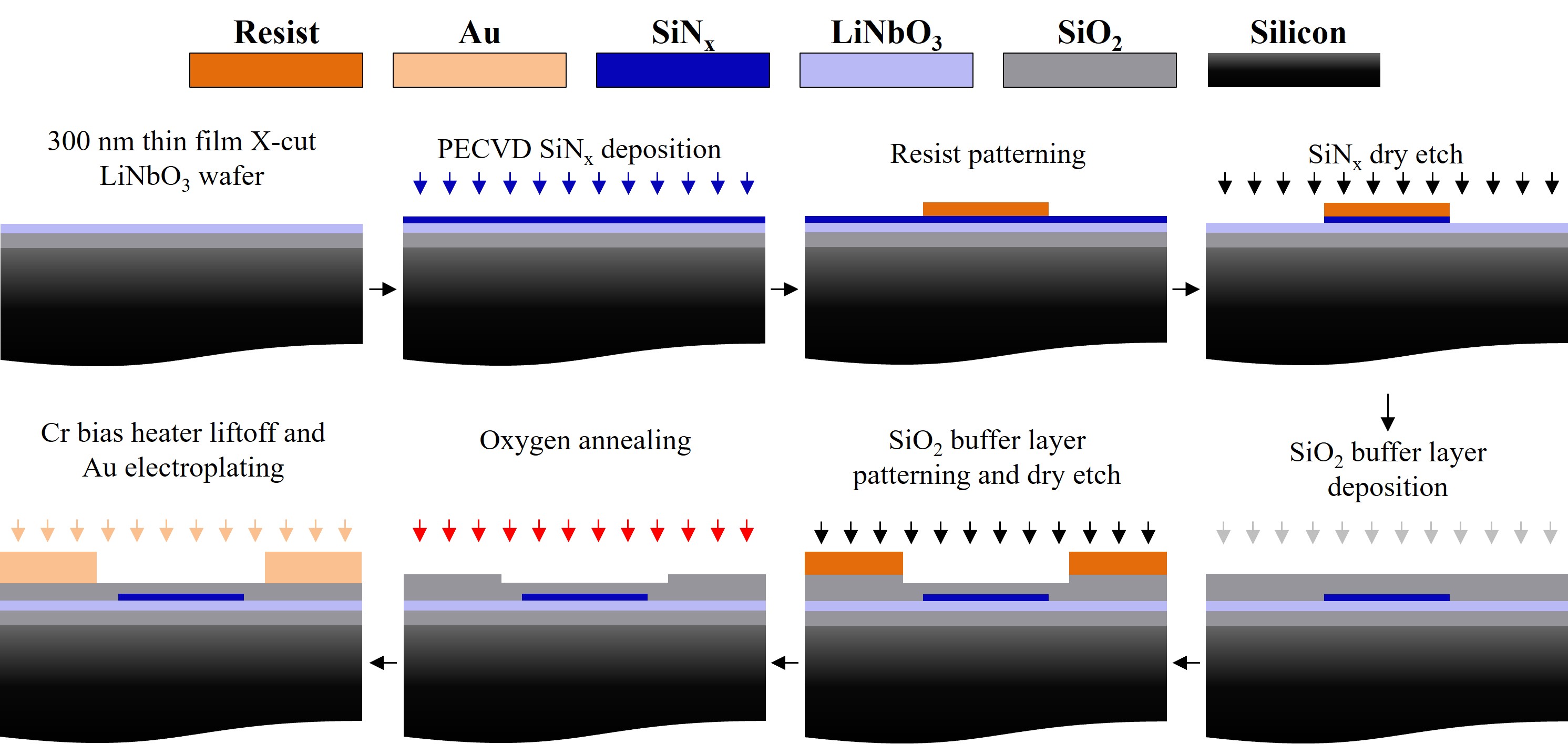}
{Simplified fabrication flow chart of the folded SiN$_\mathrm{x}$-LiNbO$_\mathrm{3}$ modulator.}
\label{fig:fabdevice}

Fabrication of the folded modulator begins on a 300 nm thick X-cut TFLNOI substrate procured from NanoLN$^\mathrm{TM}$. A 4.7 $\mu$m layer of thermal SiO$_\mathrm{2}$ below the LiNbO$_\mathrm{3}$ is bonded to a 500 $\mu$m thick Si handle. Electron-beam lithography (EBL) is used to pattern alignment structures in TiW (10$\%$/90$\%$) for future lithographic steps. A 100 nm thick layer of SiN$_\mathrm{x}$ is deposited on top of the thin-film LiNbO$_\mathrm{3}$ using PECVD. This layer will be used to define the SiN$_\mathrm{x}$ strip-loaded waveguides. The refractive index for the SiN$_\mathrm{x}$ is 1.943 at 1550 nm. EBL is again used to define the optical structures. The features are defined via dry etching of the SiN$_\mathrm{x}$ layer in an inductively coupled plasma (ICP). A 2 $\mu$m SiO$_\mathrm{2}$ buffer layer is then deposited using PECVD. A laser writer is used to pattern the regions where an increased thickness of the SiO$_\mathrm{2}$ buffer layer will protect the optical mode from interacting with an overhead Au electrode. The rest of the SiO$_\mathrm{2}$ buffer layer is reduced to 450 nm thickness via dry etching in an ICP. To reduce RF absorption losses, the device is annealed in an O$_\mathrm2$ environment. A seed layer is deposited onto the sample using an Angstrom metal sputtering system and the electrodes are defined via EBL. An electroplating process is then used to realize the 1 $\mu$m thick Au electrodes. Finally, the device's waveguide facets are polished to increase fiber coupling efficiency. A simplified flow of the fabrication steps is found in Fig \ref{fig:fabdevice}. An image of the fabricated modulator is shown in Fig. \ref{fig:solidworks}. (b). This process is expanded upon in previous work \cite{nelan_compact_2022}.

\section{Optical and RF characterization}

\subsection{Optical loss characterization}

A tunable telecom laser (Keysight 81608A) is used at a wavelength of 1550 nm and connected to an OZ Optics 1x4 lensed fiber array, with a mode field diameter (MFD) of roughly 2.5 $\mu$m through a polarization-maintaining (PM) fiber patch cable. The laser output from the lensed fiber array is free-space coupled to the on-chip fiber coupler. Due to the MFD mismatch between the lensed fiber and on-chip coupler, there is a 4.44 dB coupling loss per facet. Now, the tunable laser source is replaced with a reflectometer/light-wave analyzer (Luna OBR 6415), and the optical back-scatter and propagation loss from a reference structure is measured and shown in Fig. \ref{fig:vpidata}. (d) to be .64 dB/cm. The total insertion loss of the device is roughly 12.0 dB. The on-chip loss after the 17 mm path length is roughly 1.1 dB. This accounts for roughly 9.98 dB total insertion loss. The additional 2.02 dB loss is incurred due to the difference in extension between the lensed fibers in the lensed fiber array, preventing proper free-space fiber alignment. In future work, the on-chip fiber coupler can be made to work with a flat-polished v-groove array (VGA), to eliminate losses from fiber array alignment.

\Figure[h](topskip=0pt, botskip=0pt, midskip=0pt)[width=0.9\textwidth]{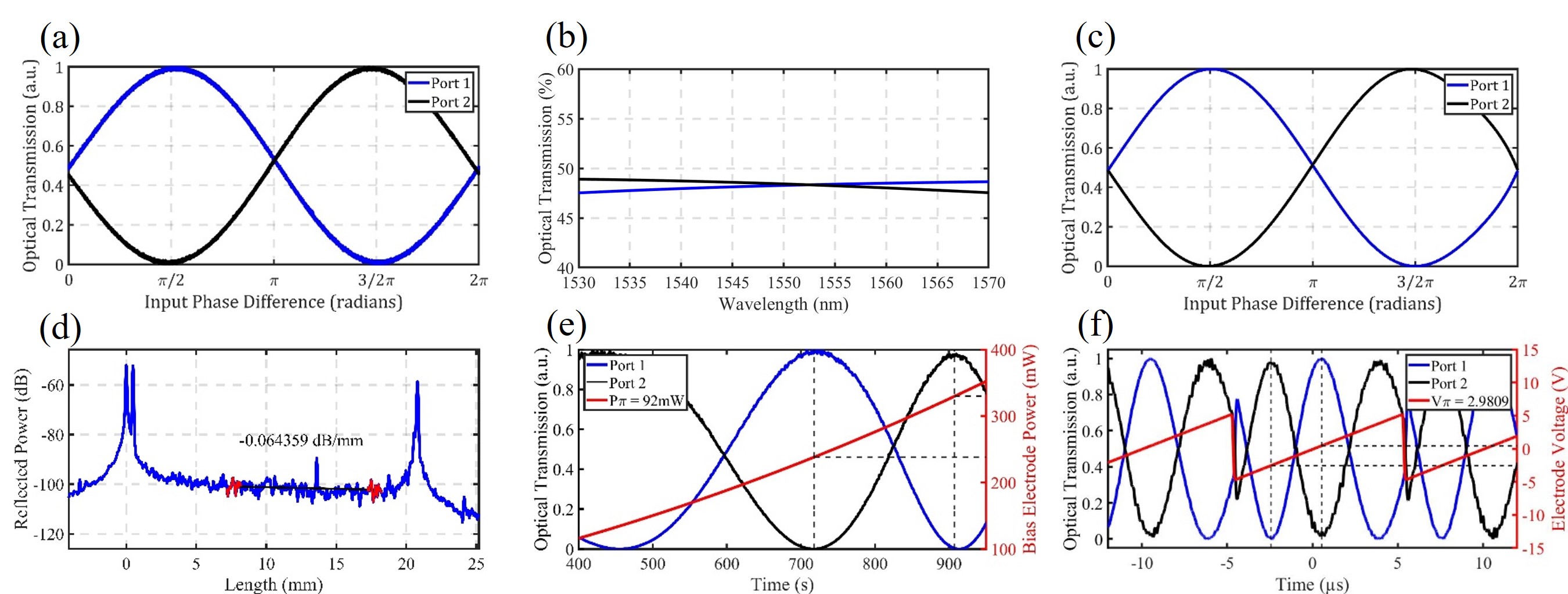}
{(a) Reference 2x2 MMI transmission in each output port vs phase difference between each input port. Data experimentally collected.
(b) Simulated 2x2 MMI transmission in each output port vs input wavelength across optical C-band. Peak splitting difference is 1.4$\%$.
(c) Reference 2x2 MMI transmission in each output port vs phase difference between each input port. Data simulated in ANSYS Lumerical FDTD.
(d) Measured propagation loss of the on-chip reference structure. Measured propagation loss is 0.64 dB/cm.
(e) Measured optical power of the folded modulator at each output port vs power applied to the thermal bias electrode. P$\pi$ = 92 mW.
(f) Measured optical power of the folded modulator at each output port vs voltage applied to the RF GSG electrodes. V$\pi$ = 2.98 V.}
\label{fig:vpidata}

\subsection{2x2 MMI input response characterization}

The 2x2 MMI must provide two arms with inversely proportional field intensities, varying with the phase difference between both input ports if it is to be used in a MZM configuration for a balanced link. Because one output of the ideal 2x2 MMI contains a $\frac{\pi}{2}$ phase shift with respect to the other when fed through one input arm, and considering the reciprocal nature of the device, illuminating both input arms with phase shift of $\pi$ enables switching of light from one port to the other port. We can use the field transfer matrix of a 2x2 MMI, shown in Eq. (2), to show the relationship between the difference in phase between the two input arms ($\Delta\phi$), and the transmission through each output arm [19]:
\begin{equation}
\begin{bmatrix} A^{'} \\ B^{'} \end{bmatrix} = 
e^{j\phi_0}\begin{bmatrix}1&&e^{\frac{j\pi}{2} }\\e^{\frac{j\pi}{2} }&&1\end{bmatrix}\begin{bmatrix}A\\B\end{bmatrix}.
\end{equation}
The power transferred through each arm is modeled using $\frac{ |A'|^{2} }{ 2 }$ and $\frac{ |B'|^{2} }{ 2 }$. In Fig. \ref{fig:vpidata}. (b), it is confirmed that the MMI can maintain an equal splitting ratio across a range of optical wavelengths, where a peak splitting difference of 1.4$\%$ is found at the 1530 nm input wavelength. This still supports an ER of over 40 dB, shown later in this work. In Fig. \ref{fig:vpidata}. (c), the same relationship is demonstrated using Lumerical’s FDTD simulation software. Here, two Gaussian mode sources are applied to the inputs of the structure, while one’s phase is swept from 0 to 2$\pi$ radians. The normalized power transmission through each output arm is plotted along this sweep.

An imbalanced Mach-Zehnder Interferometer (MZI) is constructed from a directional fiber coupler, where a tunable laser source is equally split among two fibers of lengths 1 m and 1.5 m, and then connected to a 1x2 OZ Optics v-groove array (VGA), aligned to launch TE-polarized light at 1550 nm. The VGA is aligned to the input waveguides of the 2x2 MMI. An identical VGA is aligned to the output waveguides of the device and each fiber is fed to a broadband InGaAs photodetector before being directed to an oscilloscope. The laser source is swept from 1549.9 nm to 1550.1 nm at 10 nm/s, at a step length of 0.1 nm. Because there is a length difference in the fibers after the 3 dB directional coupler and before the MMI, sweeping the laser source creates a periodic phase variation among the input ports of the MMI. The output power of each port is then captured simultaneously on the oscilloscope and can be seen in Fig. \ref{fig:vpidata}. (a). Here, an inversely proportional optical intensity between both device’s output ports is experimentally validated. The same experiment is performed in Lumerical FDTD and shown in Fig. \ref{fig:vpidata}. (c). Here, both inputs of the device are illuminated with an equal optical intensity, and an input phase sweep is performed to show that an inversely proportional relationship between the optical intensity at the output ports exists.

\subsection{DC response characterization}

The MZM functionality as an electro-optic (EO) transducer is measured at low frequencies and plotted in Fig. \ref{fig:vpidata}. (f). The observed V$\pi$ is our chosen figure-of-merit (FOM). A low speed (150 kHz) triangular voltage wave is applied to the GSG electrodes using an arbitrary waveform generator via a GSG probe. The device's electrodes are excited such that the center electrode is active, while the outer electrodes are grounded. The optical output signal from each output port is connected to a broadband photodetector, which is connected to an adjustable trans-impedance amplifier (TIA) and connected to an oscilloscope. The output of the arbitrary waveform generator and photodetectors are observed on the oscilloscope and presented in Fig. \ref{fig:vpidata}. (f). The measured V$\pi$ agrees with the simulated value of roughly 3 V.

The same experiment is then performed using a thermo-optic (TO) phase shifter near the fold of the device to ensure the device can be biased to null, quadrature and peak bias. The input power is swept from 100 mW to 350 mW and the output of both photodetectors is plotted alongside the applied TO phase shifter power in Fig. \ref{fig:vpidata}. (e). The power required to shift from null to peak bias (P$\pi$) is found to be P$\pi$ = 92 mW.

\subsection{Extinction ratio characterization}

\Figure[h](topskip=0pt, botskip=0pt, midskip=0pt)[width=0.7\textwidth]{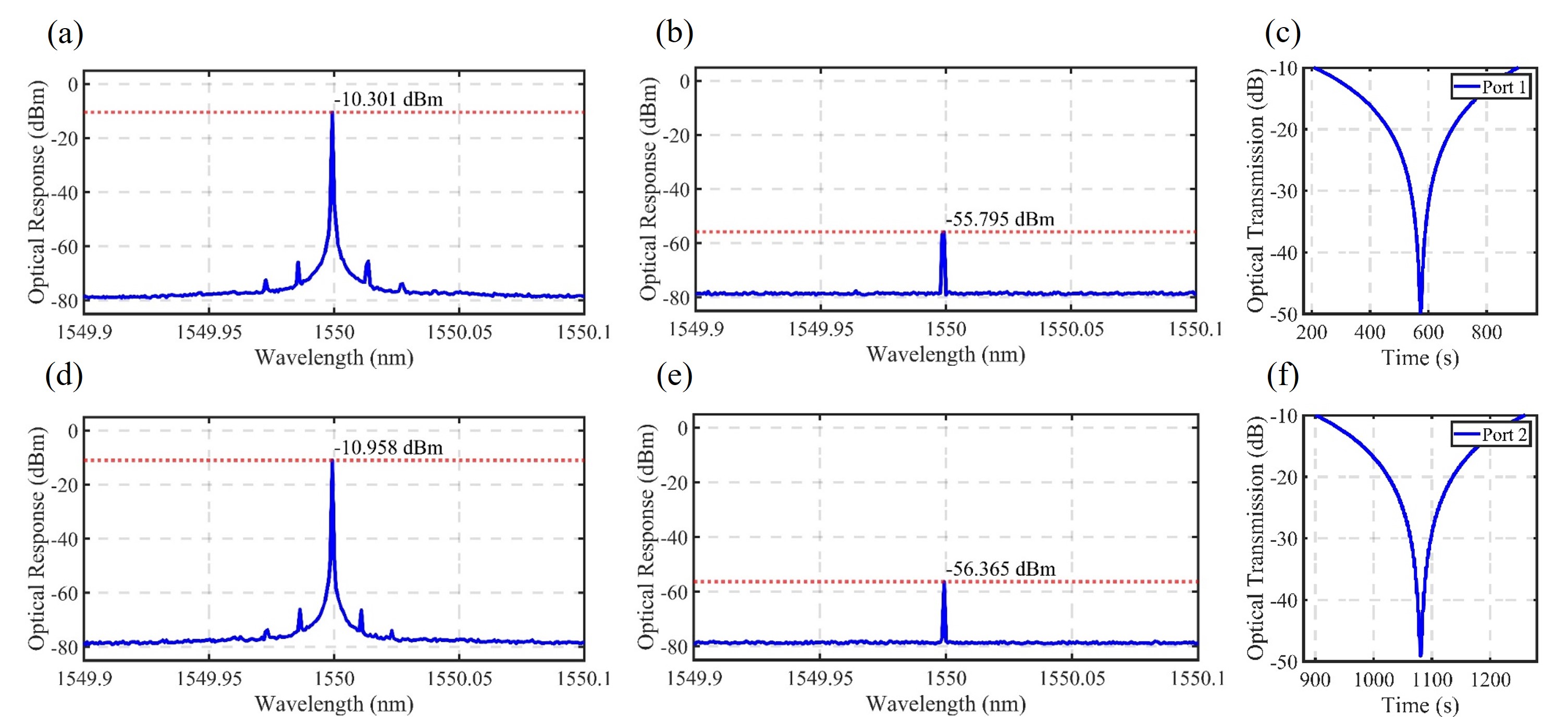}
{(a) Port 1 optical power output at respective peak bias while forward propagating through the modulator.
(b) Port 1 optical power output at respective null bias while forward propagating through the modulator.
(c) Port 1 optical power output while the bias condition is swept through respective null.
(d) Port 2 optical power output at respective peak bias while forward propagating through the modulator.
(e) Port 2 optical power output at respective null bias while forward propagating through the modulator.
(f) Port 2 optical power output while the bias condition is swept through respective null. Data collected using an APEX Optical Spectrum Analyzer and Thor Labs handheld power meter. Input power is 1 mW, input and output fibers are separated from end facets to increase measurement stability. Peak power is not representative of total insertion loss.}
\label{fig:extforward}

Through careful design of the 2x2 MMI splitter/combiner used at the output of the device, the optical extinction ratio can be improved compared to previous work which employs a 2x1 MMI combiner \cite{nelan_compact_2022}. When a 2x1 MMI is used at the output of an MZM, and the MZM is biased to a null condition such that there is a $\pi$ phase difference between both modulation arms, the light entering the 2x1 MMI will interfere destructively at the output port and create substrate modes in the LiNbO$_\mathrm{3}$ slab. While very little light may immediately enter the output waveguide, a portion of optical power from the substrate modes will couple back into the waveguide after some length, leading to a degraded extinction ratio.

To solve this, we have placed a 2x2 MMI after the interaction region to serve as the splitter/combiner at the output of the MZM. Now, moving from a peak to null bias condition will shift the light from one output port to another rather than exciting unwanted substrate modes. The addition of linear tapers on the input and output of the MMI further prevents reflections cause by MFD mismatch between the multimode region and the waveguides leading to substrate modes. Moreover, because the optical mode is primarily confined in the LiNbO$_\mathrm{3}$ slab, the multimode region of the 2x2 MMI must be relatively long compared to Silicon on Insulator (SOI) or silicon on LiNbO$_\mathrm{3}$ strip-loaded waveguides, which maintain a higher index contrast than SiN$_\mathrm{x}$ on LiNbO$_\mathrm{3}$. As such, the LiNbO$_\mathrm{3}$-SiN$_\mathrm{x}$ material system affords greater fabrication tolerance, where, for example, a +/- 25 nm variation in the length of the multimode region is only .004$\%$ of the total length.

\Figure[h](topskip=0pt, botskip=0pt, midskip=0pt)[width=0.6\textwidth]{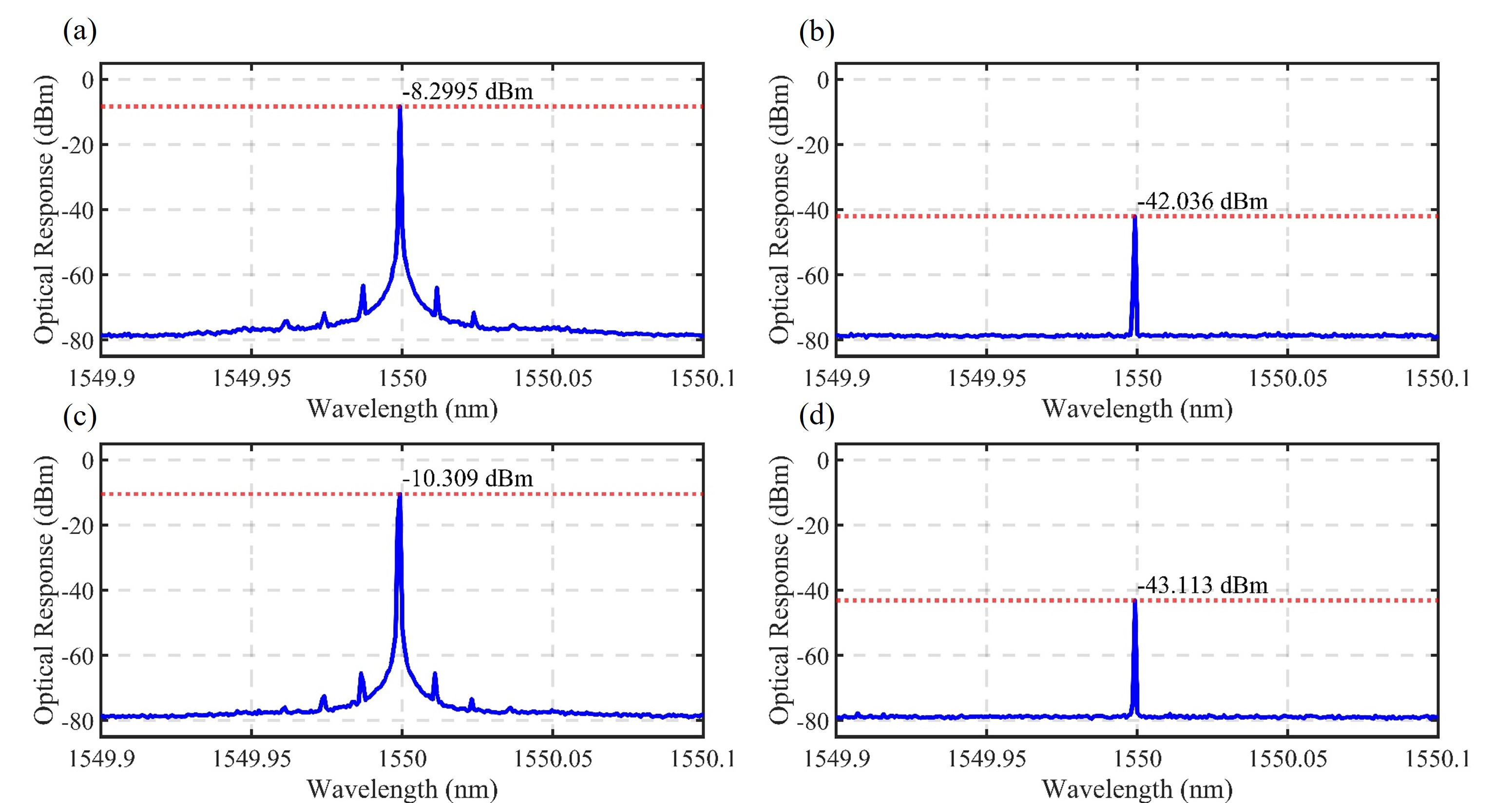}
{(a) Power output of the modulator at peak bias when fed backwards through output port 1.
(b) Power output of the modulator at null bias when fed backwards through output port 1.
(c) Power output of the modulator at peak bias when fed backwards through output port 2.
(d) Power output of the modulator at null bias when fed backwards through output port 2. Data collected using an Optical Spectrum Analyzer. Input power is 1 mW, input and output fibers are separated from end facets to increase measurement stability. Peak power is not representative of total insertion loss.}
\label{fig:extbackward}

To demonstrate the ability of our 2x2 MMI to improve the extinction ratio of the MZM, we have measured the extinction ratio of the device under forward and reverse operation. In forward operation, light enters the device through the input port as defined in Fig. \ref{fig:solidworks}. (b), is split into two paths using a 1x2 MMI, and then is recombined after passing through the interaction region using a 2x2 MMI. Thermo-optic phase shifters, shown in Fig. \ref{fig:schematic} are used to sweep the bias condition of the device, and the peak and null intensity of each output port is collected on an Optical Spectrum Analyzer (OSA). The optical response of port 1 at peak and null bias is plotted in Fig. \ref{fig:extforward}. (a) and (b), respectively. The optical response of port 2 at peak and null bias is plotted in Fig. \ref{fig:extforward}. (d) and (c), respectively. 

To validate these results, the optical transmission in is then collected using an optical power meter, and plotted to find the static extinction ratio of the device as the bias condition is swept through null bias for each output port, shown in Fig. \ref{fig:extforward}. (c) and (f), corresponding to output ports 1 and 2, respectively. Multiple devices were measured to confirm the repeatability of this result. The optical power meter shows a higher ER than the OSA. To avoid misrepresentation of data, the lower number acquired from the OSA is used.

\Figure[h](topskip=0pt, botskip=0pt, midskip=0pt)[width=0.9\columnwidth]{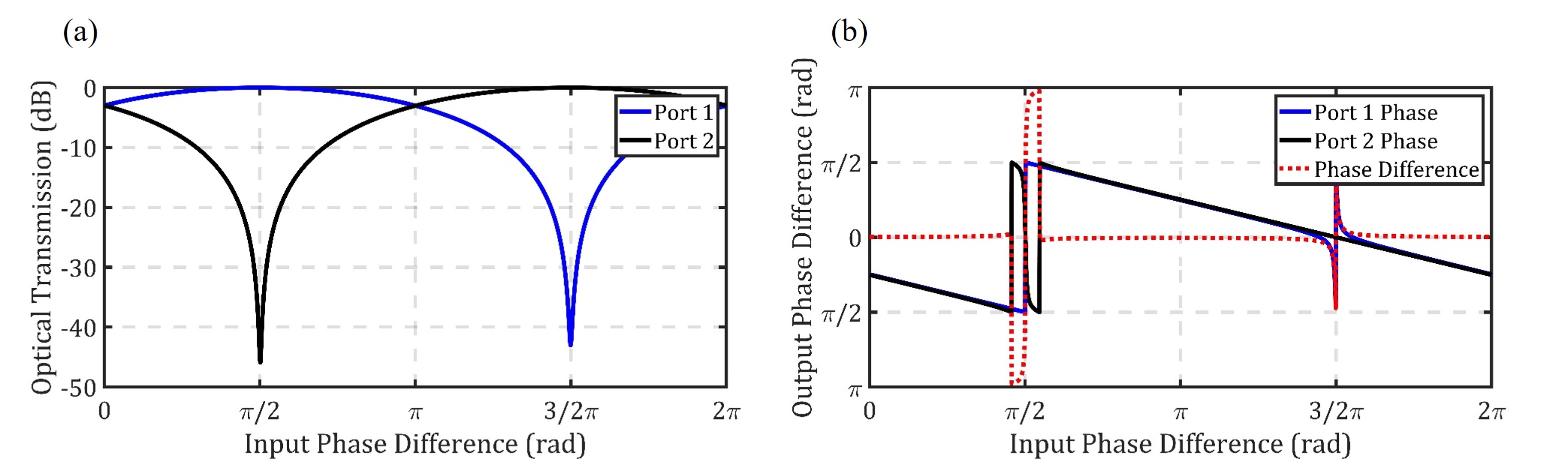}
{(a) Calculated extinction ratio from each port in the 2x2 MMI with applied splitting imbalance and input power imbalance.
(b) Calculated relative phase relation from each port in the 2x2 MMI with applied splitting imbalance and input power imbalance.}
\label{fig:calculatedER}

The same experiment is then performed with the device in reverse operation. Here, the laser source is connected to either output port 1 or 2 as defined in Fig. \ref{fig:solidworks}. (b). Now, the 1x2 MMI splitter at the input acts as a 2x1 MMI combiner\cite{wang_proposal_2014,ahmed_subvolt_2020,khalil_two-dimensional_2004}. Thermo-optic phase shifters are again used to sweep the bias condition of the device, and the peak and null intensity of the output is collected on an OSA. The extinction ratio is measured using both ports 1 and 2 as the input and plotted in Fig. \ref{fig:extbackward}. The optical response of the device when fed from port 1 is plotted in Fig. \ref{fig:extbackward}. (a) and (b), respectively. The optical response of the device when fed from port 2 is plotted in Fig. \ref{fig:extbackward}. (c) and (d), respectively.

The extinction ratio of the device in forward operation is roughly 45 dB, while the extinction ratio of the device in reverse operation is roughly 33 dB. This demonstrates the effectiveness of the 2x2 MMI in this material system, where a 45 dB extinction ratio is achieved without the need for cascaded directional couplers.

A 2x2 MMI transfer matrix is used to simulate the result after the fact to find the input power and splitting imbalance errors. The transfer matrix for this is shown below:

\begin{equation}
\begin{bmatrix}
A' \\
B'
\end{bmatrix} =
e^{j \phi_0}
\begin{bmatrix}
1 & e^{j \frac{\pi}{2}(\phi_{imbal})} \\
e^{j \frac{\pi}{2}(\phi_{imbal})} & 1
\end{bmatrix}
\begin{bmatrix}
Ae^{j \phi_A} \\
Be^{j \phi_B}
\end{bmatrix}
\begin{bmatrix}
P_{imbal} \\
1
\end{bmatrix}.
\end{equation}
Here, $A$ and $B$ are the input power of both input ports while $A'$ and $B'$ represent the output power of the MMI output ports. The term $\phi_0$ is the constant phase offset that the MMI imparts on the optical mode. The $\pi/2$ term accounts for the ideal $\pi/2$ phase offset between the output ports when the MMI is fed by one input port. The term $\phi_{imbal}$ accounts for any discrepancies between the ideal $\pi/2$ phase offset and the actual phase offset, which may be different due to fabrication tolerances or design issues. Similarly, the term $P_{imbal}$ accounts for unequal splitting within the MMI. With this, the output power of output ports $A'$ and $B'$, is
\begin{equation}
A' = e^{j \phi_0} \left (P_{imbal} A e^{j \phi_A} + B e^{j \phi_B} e^{j \frac{\pi}{2} \phi_{imbal}}\right )
\end{equation}
\begin{equation}
B' = e^{j \phi_0} \left (P_{imbal} A e^{j \phi_A}e^{j \frac{\pi}{2} \phi_{imbal}} + B e^{j \phi_B}\right ).
\end{equation}
Using values $P_{imbal}$ = 0.998, $\phi_{imbal}$ = 0.998, $A$ = 1, $B$ = 0.988, the ER and phase relations of the 2x2 MMI are calculated and plotted in Fig. \ref{fig:calculatedER}. (a) and (b), respectively. It is shown that this aligns well with the measured data and demonstrates that the 2x2 MMI will still produce a high ER when input power and splitting ratio are not perfectly balanced. If fabrication issues are expected, an additional thermal heater can be placed on the top or bottom of the MMI to tune the optical splitting ratio and account for fabrication errors \cite{Rosa:16}. Finally, the MMI can maintain a over 40 dB ER across the optical C-band, where a peak splitting difference of 1.4$\%$ is found at 1530 nm. 

\subsection{High frequency EO response}

\Figure[h](topskip=0pt, botskip=0pt, midskip=0pt)[width=.95\columnwidth]{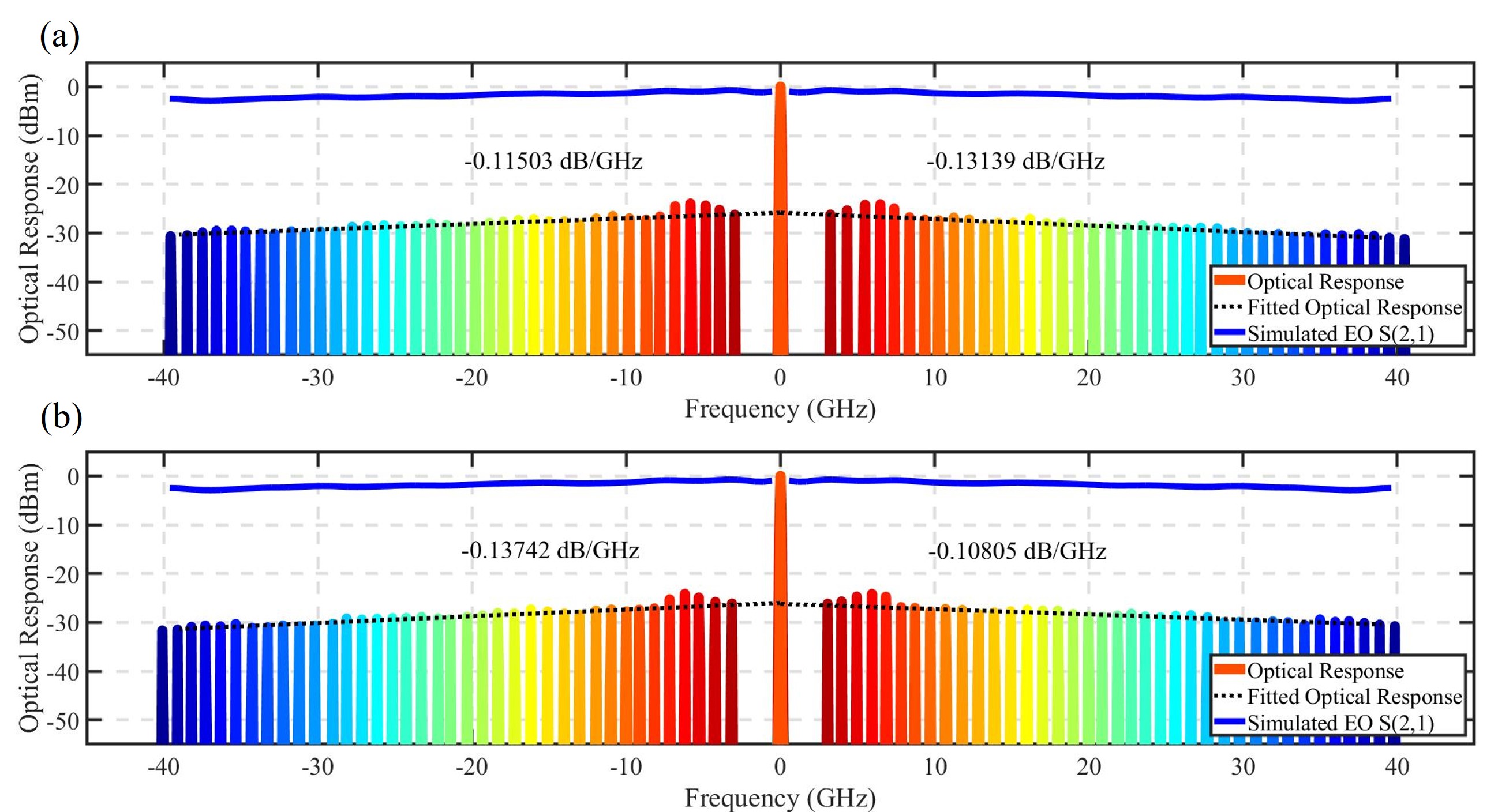}
{(a) Measured optical sidebands and simulated EO response at quadrature bias from 0 to 40 GHz using a 1550 nm wavelength carrier signal. Measurement taken at output port 1, UV15 index matching fluid present.
(b) Measured optical sidebands and simulated EO response at quadrature bias from 0 to 40 GHz using a 1550 nm wavelength carrier signal. Measurement taken at output port 2, UV15 index matching fluid present.}
\label{fig:sidebands}

The traveling-wave electro-optic response of the MZM is achieved by measuring the optical sideband power with an RF modulating signal applied to the electrodes. A tunable telecom laser is used to launch 1550nm TE-polarized light into the device, while an RF signal generated by a signal generator is connected to the input side of the GSG electrode through a GSG probe so the RF and optical signals are co-propagating. In this way, the overlap between the modulating RF field and the modulated optical signal is maintained for as long as possible. The modulator is biased to quadrature using a DC current source connected to the TO phase shifters to ensure the modulator is operating in the most linear region of the response curve. The RF modulating signal is swept from 1 to 40 GHz in 1 GHz increments, and the intensity of the sidebands of the modulated optical signal is observed using an OSA \cite{Shi:03}. The RF source is then connected to an RF thermocouple power meter and again swept from 1 to 40 GHz to gather the RF input power at each frequency step. The modulation spectrum is normalized to the optical carrier input and RF power. The modulation spectrum of port 1 and 2 from 3 to 40 GHz at quadrature bias of the UV15-clad device is shown in Fig. \ref{fig:sidebands}. (a) and (b), respectively. Lower frequencies are omitted due to an unlevel power output from the RF source. The sideband power roll-off from low frequencies to 40 GHz is roughly 0.1 dB/GHz, excluding spurious low-frequency response, the 3 dB bandwidth is roughly 30 GHz. The experimental test setup is shown in Fig. \ref{fig:schematic}. The measured EO response is a close match to the simulated EO response extracted from the measured electrical scattering matrix \cite{Zhu:21}. This confirms that the expected optical group index is accurate, where if it was not, there would be a discrepancy between the simulated and measured result.

\Figure[h](topskip=0pt, botskip=0pt, midskip=0pt)[width=0.9\columnwidth]{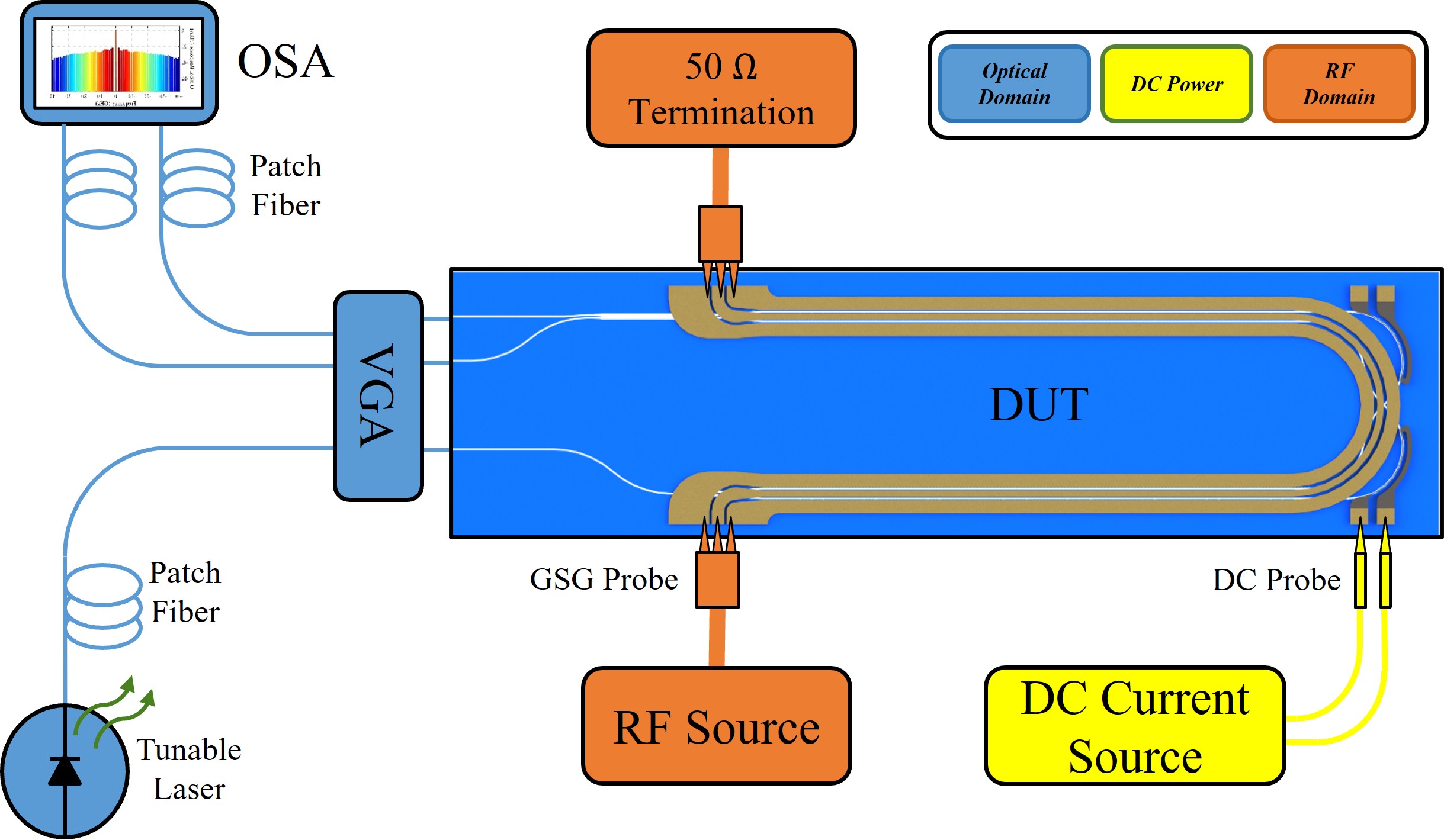}
{Schematic of the experimental setup used to collect sideband data of the device.}
\label{fig:schematic}

\section{Conclusion}

In this work, a high extinction ratio hybrid LiNbO$_\mathrm{3}$-SiN$_\mathrm{x}$ electro-optic MZM with two equal and opposite outputs is designed, fabricated and characterized. The device uses a 2x2 MMI at the output to provide two inversely proportional outputs with exceptional extinction ratio performance. Fiber couplers are used at the input and output to lower the total insertion loss to 12 dB. To the best of our knowledge, this is the first time a high extinction ratio, folded, dual output LiNbO$_\mathrm{3}$-SiN$_\mathrm{x}$ electro-optic dual-output MZM with a waveguide crossing has been demonstrated in this material platform. The device presented shows a measured DC-V$\pi$ of 3.0 V with an 11 mm interaction region, and an extinction ratio of over 45 dB. This design does not require cascaded directional couplers to control splitting imbalance, and shows repeatability across multiple devices. This is nearly 31 times higher than high-extinction ratio single-output modulators in literature today. There is a clear path forward to optimizing either or both the low and high frequency operation depending on application-specific requirements. In future work, capacitively-loaded traveling-wave electrodes will be incorporated to increase the operational bandwidth.

A dual-output MZM can be paired with a balanced PD receiver and used in many photonic communication systems where a high extinction, dual output device is paramount to the high-performance operation of the system as a whole. This enables heavy, lossy and space-consuming cables to be replaced with much lighter optical fibers. A balanced PD receiver pair is capable of detecting both the signal amplitude and phase information and is compatible with complex modulation formats used in DPSK and DQPSK systems. Moreover, the demonstration of the 2x2 MMI coupler to reach >45 dB ER lends itself to exotic applications such as IR and mid-IR Bracewell nulling interferometry for exoplanet detection \cite{Goldsmith:17}.

\bibliographystyle{unsrt}
\bibliography{ReferencesDual}

\begin{thebibliography}{10}

\bibitem{alvarado_impact_2016}
Alex Alvarado, David~J. Ives, Seb~J. Savory, and Polina Bayvel.
\newblock On the {Impact} of {Optimal} {Modulation} and {FEC} {Overhead} on
  {Future} {Optical} {Networks}.
\newblock {\em Journal of Lightwave Technology}, 34(9):2339--2352, May 2016.

\bibitem{winzer_scaling_2017}
Peter~J. Winzer and David~T. Neilson.
\newblock From {Scaling} {Disparities} to {Integrated} {Parallelism}: {A}
  {Decathlon} for a {Decade}.
\newblock {\em Journal of Lightwave Technology}, 35(5):1099--1115, March 2017.

\bibitem{cheng_recent_2018}
Qixiang Cheng, Meisam Bahadori, Madeleine Glick, Sébastien Rumley, and Keren
  Bergman.
\newblock Recent advances in optical technologies for data centers: a review.
\newblock {\em Optica}, 5(11):1354, November 2018.

\bibitem{yue_experimental_2019}
Yang Yue, Qiang Wang, and Jon Anderson.
\newblock Experimental {Investigation} of 400 {Gb}/s {Data} {Center}
  {Interconnect} {Using} {Unamplified} {High}-{Baud}-{Rate} and {High}-{Order}
  {QAM} {Single}-{Carrier} {Signal}.
\newblock {\em Applied Sciences}, 9(12):2455, June 2019.

\bibitem{wooten_review_2000}
E.L. Wooten, K.M. Kissa, A.~Yi-Yan, E.J. Murphy, D.A. Lafaw, P.F. Hallemeier,
  D.~Maack, D.V. Attanasio, D.J. Fritz, G.J. McBrien, and D.E. Bossi.
\newblock A review of lithium niobate modulators for fiber-optic communications
  systems.
\newblock {\em IEEE Journal of Selected Topics in Quantum Electronics},
  6(1):69--82, January 2000.

\bibitem{carey_millimeter_2021}
Victoria~A. Carey, Matthew~R. Konkol, Shouyuan Shi, Andrew~J. Mercante, Kevin
  Shreve, Andrew~A. Wright, Christopher~A. Schuetz, and Dennis~W. Prather.
\newblock Millimeter {Wave} {Photonic} {Tightly} {Coupled} {Array}.
\newblock {\em IEEE Transactions on Antennas and Propagation},
  69(8):4488--4503, August 2021.

\bibitem{beardell_rf-photonic_2021}
William Beardell, Benjamin Mazur, Conor~J Ryan, Garrett~J. Schneider, Janusz
  Murakowski, and Dennis~W. Prather.
\newblock {RF}-{Photonic} {Spatial}-{Spectral} {Channelizing} {Receiver}.
\newblock {\em Journal of Lightwave Technology}, pages 1--1, 2021.

\bibitem{nelan_compact_2022}
Sean Nelan, Andrew Mercante, Cooper Hurley, Shouyuan Shi, Peng Yao, Benjamin
  Shopp, and Dennis~W. Prather.
\newblock Compact thin film lithium niobate folded intensity modulator using a
  waveguide crossing.
\newblock {\em Optics Express}, 30(6):9193, March 2022.

\bibitem{liu_low_2021}
Ye~Liu, Heng Li, Jia Liu, Su~Tan, Qiaoyin Lu, and Weihua Guo.
\newblock Low {V} $_{\textrm{pi}}$ thin-film lithium niobate modulator
  fabricated with photolithography.
\newblock {\em Optics Express}, 29(5):6320, March 2021.

\bibitem{chen_high_2022}
Gengxin Chen, Kaixuan Chen, Ranfeng Gan, Ziliang Ruan, Zong Wang, Pucheng
  Huang, Chao Lu, Alan Pak~Tao Lau, Daoxin Dai, Changjian Guo, and Liu Liu.
\newblock High performance thin-film lithium niobate modulator on a silicon
  substrate using periodic capacitively loaded traveling-wave electrode.
\newblock {\em APL Photonics}, 7(2):026103, February 2022.

\bibitem{zhang_high_2016}
Xingyu Zhang, Chi-Jui Chung, Amir Hosseini, Harish Subbaraman, Jingdong Luo,
  Alex K-Y. Jen, Robert~L. Nelson, Charles Y-C. Lee, and Ray~T. Chen.
\newblock High {Performance} {Optical} {Modulator} {Based} on {Electro}-{Optic}
  {Polymer} {Filled} {Silicon} {Slot} {Photonic} {Crystal} {Waveguide}.
\newblock {\em Journal of Lightwave Technology}, 34(12):2941--2951, June 2016.

\bibitem{zhang_integrated_2021}
Mian Zhang, Cheng Wang, Prashanta Kharel, Di~Zhu, and Marko Lončar.
\newblock Integrated lithium niobate electro-optic modulators: when performance
  meets scalability.
\newblock {\em Optica}, 8(5):652, May 2021.

\bibitem{he_high-performance_2019}
Mingbo He, Mengyue Xu, Yuxuan Ren, Jian Jian, Ziliang Ruan, Yongsheng Xu,
  Shengqian Gao, Shihao Sun, Xueqin Wen, Lidan Zhou, Lin Liu, Changjian Guo,
  Hui Chen, Siyuan Yu, Liu Liu, and Xinlun Cai.
\newblock High-performance hybrid silicon and lithium niobate
  {Mach}–{Zehnder} modulators for 100 {Gbit} s-1 and beyond.
\newblock {\em Nature Photonics}, 13(5):359--364, May 2019.

\bibitem{ahmed_high-efficiency_2020}
Abu Naim~R. Ahmed, Shouyuan Shi, Andrew Mercante, Sean Nelan, Peng Yao, and
  Dennis~W. Prather.
\newblock High-efficiency lithium niobate modulator for {K} band operation.
\newblock {\em APL Photonics}, 5(9):091302, September 2020.

\bibitem{mercante_thin_2018}
Andrew~J. Mercante, Shouyuan Shi, Peng Yao, Linli Xie, Robert~M. Weikle, and
  Dennis~W. Prather.
\newblock Thin film lithium niobate electro-optic modulator with terahertz
  operating bandwidth.
\newblock {\em Optics Express}, 26(11):14810, May 2018.

\bibitem{rao_compact_2018}
Ashutosh Rao and Sasan Fathpour.
\newblock Compact {Lithium} {Niobate} {Electrooptic} {Modulators}.
\newblock {\em IEEE Journal of Selected Topics in Quantum Electronics},
  24(4):1--14, July 2018.

\bibitem{reed_silicon_2010}
G.~T. Reed, G.~Mashanovich, F.~Y. Gardes, and D.~J. Thomson.
\newblock Silicon optical modulators.
\newblock {\em Nature Photonics}, 4(8):518--526, August 2010.

\bibitem{weigel_bonded_2018}
Peter~O. Weigel, Jie Zhao, Kelvin Fang, Hasan Al-Rubaye, Douglas Trotter, Dana
  Hood, John Mudrick, Christina Dallo, Andrew~T. Pomerene, Andrew~L. Starbuck,
  Christopher~T. DeRose, Anthony~L. Lentine, Gabriel Rebeiz, and Shayan
  Mookherjea.
\newblock Bonded thin film lithium niobate modulator on a silicon photonics
  platform exceeding 100 {GHz} 3-{dB} electrical modulation bandwidth.
\newblock {\em Optics Express}, 26(18):23728, September 2018.

\bibitem{horst_cascaded_2013}
Folkert Horst, William~M.J. Green, Solomon Assefa, Steven~M. Shank, Yurii~A.
  Vlasov, and Bert~Jan Offrein.
\newblock Cascaded {Mach}-{Zehnder} wavelength filters in silicon photonics for
  low loss and flat pass-band {WDM} (de-)multiplexing.
\newblock {\em Optics Express}, 21(10):11652, May 2013.

\bibitem{huo_diamond_2016}
Dehong Huo, Zi~Jie Choong, Yilun Shi, John Hedley, and Yan Zhao.
\newblock Diamond micro-milling of lithium niobate for sensing applications.
\newblock {\em Journal of Micromechanics and Microengineering}, 26(9):095005,
  September 2016.

\bibitem{wang_integrated_2018}
Cheng Wang, Mian Zhang, Xi~Chen, Maxime Bertrand, Amirhassan Shams-Ansari,
  Sethumadhavan Chandrasekhar, Peter Winzer, and Marko Lončar.
\newblock Integrated lithium niobate electro-optic modulators operating at
  {CMOS}-compatible voltages.
\newblock {\em Nature}, 562(7725):101--104, October 2018.

\bibitem{alloatti_100_2014}
Luca Alloatti, Robert Palmer, Sebastian Diebold, Kai~Philipp Pahl, Baoquan
  Chen, Raluca Dinu, Maryse Fournier, Jean-Marc Fedeli, Thomas Zwick, Wolfgang
  Freude, Christian Koos, and Juerg Leuthold.
\newblock 100 {GHz} silicon–organic hybrid modulator.
\newblock {\em Light: Science \& Applications}, 3(5):e173--e173, May 2014.

\bibitem{haffner_all-plasmonic_2015}
C.~Haffner, W.~Heni, Y.~Fedoryshyn, J.~Niegemann, A.~Melikyan, D.~L. Elder,
  B.~Baeuerle, Y.~Salamin, A.~Josten, U.~Koch, C.~Hoessbacher, F.~Ducry,
  L.~Juchli, A.~Emboras, D.~Hillerkuss, M.~Kohl, L.~R. Dalton, C.~Hafner, and
  J.~Leuthold.
\newblock All-plasmonic {Mach}–{Zehnder} modulator enabling optical
  high-speed communication at the microscale.
\newblock {\em Nature Photonics}, 9(8):525--528, August 2015.

\bibitem{coward_photonic_1993}
J.F. Coward, C.H. Chalfant, and P.H. Chang.
\newblock A photonic integrated-optic {RF} phase shifter for phased array
  antenna beam-forming applications.
\newblock {\em Journal of Lightwave Technology}, 11(12):2201--2205, December
  1993.

\bibitem{ahmed_subvolt_2020}
Abu Naim~R. Ahmed, Sean Nelan, Shouyuan Shi, Peng Yao, Andrew Mercante, and
  Dennis~W. Prather.
\newblock Subvolt electro-optical modulator on thin-film lithium niobate and
  silicon nitride hybrid platform.
\newblock {\em Optics Letters}, 45(5):1112, March 2020.

\bibitem{jin_high-extinction_2019}
Mingwei Jin, Jia-Yang Chen, Yong~Meng Sua, and Yu-Ping Huang.
\newblock High-extinction electro-optic modulation on lithium niobate thin
  film.
\newblock {\em Optics Letters}, 44(5):1265, March 2019.

\bibitem{wang_thin-film_2022}
Xuanhao Wang, Chenglin Shang, An~Pan, Xingran Cheng, Tao Gui, Shuai Yuan,
  Chengcheng Gui, Keshuang Zheng, Peijie Zhang, Xiaolu Song, Yanbo Li,
  Liangchuan Li, Cheng Zeng, and Jinsong Xia.
\newblock Thin-film lithium niobate based dual-polarization {IQ} modulator for
  single-carrier 1.{6Tb}/s transmission.
\newblock In Sonia~M. García-Blanco and Pavel Cheben, editors, {\em Integrated
  {Optics}: {Devices}, {Materials}, and {Technologies} {XXVI}}, page~77, San
  Francisco, United States, March 2022. SPIE.

\bibitem{subbaraman_recent_2015}
Harish Subbaraman, Xiaochuan Xu, Amir Hosseini, Xingyu Zhang, Yang Zhang, David
  Kwong, and Ray~T. Chen.
\newblock Recent advances in silicon-based passive and active optical
  interconnects.
\newblock {\em Optics Express}, 23(3):2487, February 2015.

\bibitem{azadeh_low_2015}
Saeed~Sharif Azadeh, Florian Merget, Sebastian Romero-García, Alvaro
  Moscoso-Mártir, Nils von~den Driesch, Juliana Müller, Siegfried Mantl, Dan
  Buca, and Jeremy Witzens.
\newblock Low {V$\pi$} {Silicon} photonics modulators with highly linear
  epitaxially grown phase shifters.
\newblock {\em Optics Express}, 23(18):23526, September 2015.

\bibitem{dinu_third-order_2003}
M.~Dinu, F.~Quochi, and H.~Garcia.
\newblock Third-order nonlinearities in silicon at telecom wavelengths.
\newblock {\em Applied Physics Letters}, 82(18):2954--2956, May 2003.

\bibitem{min-cheol_oh_recent_2001}
{Min-Cheol Oh}, {Hua Zhang}, {Cheng Zhang}, H.~Erlig, {Yian Chang}, B.~Tsap,
  D.~Chang, A.~Szep, W.H. Steier, H.R. Fetterman, and L.R. Dalton.
\newblock Recent advances in electrooptic polymer modulators incorporating
  highly nonlinear chromophore.
\newblock {\em IEEE Journal of Selected Topics in Quantum Electronics},
  7(5):826--835, October 2001.

\bibitem{wang_nanophotonic_2018}
Cheng Wang, Mian Zhang, Brian Stern, Michal Lipson, and Marko Lončar.
\newblock Nanophotonic lithium niobate electro-optic modulators.
\newblock {\em Optics Express}, 26(2):1547, January 2018.

\bibitem{rao_high-performance_2016}
Ashutosh Rao, Aniket Patil, Payam Rabiei, Amirmahdi Honardoost, Richard
  DeSalvo, Arthur Paolella, and Sasan Fathpour.
\newblock High-performance and linear thin-film lithium niobate
  {Mach}–{Zehnder} modulators on silicon up to 50 {GHz}.
\newblock {\em Optics Letters}, 41(24):5700, December 2016.

\bibitem{huffman_integrated_2018}
Taran~Arthur Huffman, Grant~M. Brodnik, Catia Pinho, Sarat Gundavarapu, Douglas
  Baney, and Daniel~J. Blumenthal.
\newblock Integrated {Resonators} in an {Ultralow} {Loss} {Si}
  $_{\textrm{3}}$\$ {N} $_{\textrm{4}}$\$ /{SiO} $_{\textrm{2}}$\$ {Platform}
  for {Multifunction} {Applications}.
\newblock {\em IEEE Journal of Selected Topics in Quantum Electronics},
  24(4):1--9, July 2018.

\bibitem{jin_linbo_2016}
Shilei Jin, Longtao Xu, Haihua Zhang, and Yifei Li.
\newblock {LiNbO} $_{\textrm{3}}$\$ {Thin}-{Film} {Modulators} {Using}
  {Silicon} {Nitride} {Surface} {Ridge} {Waveguides}.
\newblock {\em IEEE Photonics Technology Letters}, 28(7):736--739, April 2016.

\bibitem{stocchi_mid-infrared_2019}
M.~Stocchi, D.~Mencarelli, L.~Pierantoni, D.~Kot, M.~Lisker, A.~Göritz,
  C.~Baristiran~Kaynak, M.~Wietstruck, and M.~Kaynak.
\newblock Mid-infrared optical characterization of thin {SiN} $_{\textrm{x}}$\$
  membranes.
\newblock {\em Applied Optics}, 58(19):5233, July 2019.

\bibitem{kaloyeros_reviewsilicon_2017}
Alain~E. Kaloyeros, Fernando~A. Jové, Jonathan Goff, and Barry Arkles.
\newblock Review—{Silicon} {Nitride} and {Silicon} {Nitride}-{Rich} {Thin}
  {Film} {Technologies}: {Trends} in {Deposition} {Techniques} and {Related}
  {Applications}.
\newblock {\em ECS Journal of Solid State Science and Technology},
  6(10):P691--P714, 2017.

\bibitem{sun_folded_2021}
Shihao Sun, Mengyue Xu, Mingbo He, Shengqian Gao, Xian Zhang, Lidan Zhou, Lin
  Liu, Siyuan Yu, and Xinlun Cai.
\newblock Folded {Heterogeneous} {Silicon} and {Lithium} {Niobate}
  {Mach}–{Zehnder} {Modulators} with {Low} {Drive} {Voltage}.
\newblock {\em Micromachines}, 12(7):823, July 2021.

\bibitem{hu_folded_2021}
Jinyao Hu, Chijun Li, Changjian Guo, Chao Lu, Alan Pak~Tao Lau, Pengxin Chen,
  and Liu Liu.
\newblock Folded thin-film lithium niobate modulator based on a poled
  {Mach}–{Zehnder} interferometer structure.
\newblock {\em Opt. Lett.}, 46(12):2940--2943, June 2021.
\newblock Publisher: OSA.

\bibitem{zhang_using_2019}
Fang Zhang, Yuefeng Qi, and Wei Li.
\newblock Using optical differential phase-shift keying to solve the bipolarity
  problem of spreading code in optical time domain reflectometer.
\newblock {\em Results in Physics}, 13:102096, June 2019.

\bibitem{ren_integrated_2019}
Tianhao Ren, Mian Zhang, Cheng Wang, Linbo Shao, Christian Reimer, Yong Zhang,
  Oliver King, Ronald Esman, Thomas Cullen, and Marko Loncar.
\newblock An integrated low-voltage broadband lithium niobate phase modulator.
\newblock {\em IEEE Photonics Technology Letters}, 31(11):889--892, June 2019.

\bibitem{jin_balanced_2015}
Xiaoli Jin, Jing Su, Yaohui Zheng, Chaoyong Chen, Wenzhe Wang, and Kunchi Peng.
\newblock Balanced homodyne detection with high common mode rejection ratio
  based on parameter compensation of two arbitrary photodiodes.
\newblock {\em Optics Express}, 23(18):23859, September 2015.

\bibitem{khalil_two-dimensional_2004}
Diaa Khalil and Ayman Yehia.
\newblock Two-dimensional multimode interference in integrated optical
  structures.
\newblock {\em Journal of Optics A: Pure and Applied Optics}, 6(1):137--145,
  January 2004.

\bibitem{bachmann_general_1994}
M.~Bachmann, P.~A. Besse, and H.~Melchior.
\newblock General self-imaging properties in {N} × {N} multimode interference
  couplers including phase relations.
\newblock {\em Appl. Opt.}, 33(18):3905--3911, June 1994.
\newblock Publisher: OSA.

\bibitem{wang_proposal_2014}
Jing Wang, Minghao Qi, Yi~Xuan, Haiyang Huang, You Li, Ming Li, Xin Chen,
  Qi~Jia, Zhen Sheng, Aimin Wu, Wei Li, Xi~Wang, Shichang Zou, and Fuwan Gan.
\newblock Proposal for fabrication-tolerant {SOI} polarization splitter-rotator
  based on cascaded {MMI} couplers and an assisted bi-level taper.
\newblock {\em Optics Express}, 22(23):27869, November 2014.

\bibitem{ibarra_fuste_bandwidthlength_2013}
Jose~A. Ibarra~Fuste and Maria~C. Santos~Blanco.
\newblock Bandwidth–length trade-off figures of merit for electro-optic
  traveling wave modulators.
\newblock {\em Optics Letters}, 38(9):1548, May 2013.

\bibitem{Rosa:16}
\'{A}lvaro Rosa, Ana Guti\'{e}rrez, Antoine Brimont, Amadeu Griol, and Pablo
  Sanchis.
\newblock High performace silicon 2x2 optical switch based on a
  thermo-optically tunable multimode interference coupler and efficient
  electrodes.
\newblock {\em Opt. Express}, 24(1):191--198, Jan 2016.

\bibitem{Shi:03}
Yongqiang Shi, Lianshan Yan, and Alan~Eli Willner.
\newblock High-speed electrooptic modulator characterization using optical
  spectrum analysis.
\newblock {\em J. Lightwave Technol.}, 21(10):2358, Oct 2003.

\bibitem{Zhu:21}
Di~Zhu, Linbo Shao, Mengjie Yu, Rebecca Cheng, Boris Desiatov, C.~J. Xin,
  Yaowen Hu, Jeffrey Holzgrafe, Soumya Ghosh, Amirhassan Shams-Ansari, Eric
  Puma, Neil Sinclair, Christian Reimer, Mian Zhang, and Marko Lon\v{c}ar.
\newblock Integrated photonics on thin-film lithium niobate.
\newblock {\em Adv. Opt. Photon.}, 13(2):242--352, Jun 2021.

\bibitem{Goldsmith:17}
Harry-Dean~Kenchington Goldsmith, Nick Cvetojevic, Michael Ireland, and Stephen
  Madden.
\newblock Fabrication tolerant chalcogenide mid-infrared multimode interference
  coupler design with applications for bracewell nulling interferometry.
\newblock {\em Opt. Express}, 25(4):3038--3051, Feb 2017.

\end{thebibliography}

\EOD

\end{document}